\documentclass[a4paper,12pt,nofootinbib]{article}
\pdfoutput=1 

\usepackage{jheppub}
\usepackage{color} 
\usepackage{hyperref}
\usepackage{float}
\usepackage{ulem}
\usepackage{slashed}
\usepackage{dsfont}
\usepackage{subcaption}
\usepackage{amsmath,graphicx,amssymb,bbm}
\usepackage[all]{xy}

\begin{document}

\title{Nucleon spectra and wave functions from holographic models with dual Einstein-dilaton and Starobinsky-dilaton gravities}

\author[a]{Ad\~ao S. da Silva Junior,}
\author[b]{Juan M. Z. Pretel,}
\author[a]{and Henrique Boschi-Filho}
\affiliation[a]{Instituto de F\'{i}sica, Universidade
Federal do Rio de Janeiro, \\
Caixa Postal 68528, RJ 21941-972, Brazil.}   
\affiliation[b]{Centro Brasileiro de Pesquisas F{\'i}sicas, Rua Dr.~Xavier Sigaud, \\
150 URCA, Rio de Janeiro CEP 22290-180, RJ, Brazil}
\emailAdd{adaofisic@gmail.com}
\emailAdd{juan04manuel91@gmail.com}
\emailAdd{boschi@if.ufrj.br}

\abstract{We study the nucleon spectra in two holographic set-ups: Einstein-dilaton and Starobinsky-dilaton gravity models. The Einstein-dilaton holographic model, also known as improved holographic QCD, have been proposed some time ago to describe confinement and glueball spectra. Recently, it has been applied to the case of mesons and nucleons. In this work, we reconsider the Einstein-dilaton holographic model to discuss the nucleon spectra introducing new parameters that allow us to improve the comparison with experimental data. Then, we extend this idea to the context of Starobinsky-dilaton gravity defining another improved holographic model. We use this new holographic model to reanalyse the nucleon spectra and also compare them with soft-wall model and experimental data.}

\maketitle

\flushbottom

\section{Introduction}

Nucleons are the fundamental constituents of nuclei and account for most of the mass of visible matter of the universe. They are formed by three valence quarks plus a sea of virtual quarks and gluons, whose interaction is governed by  quantum chromodynamics (QCD), the gauge theory of the strong interaction in the Standard Model. One of the most striking properties of QCD is asymptotic freedom \cite{Gross:1973id, Politzer:1973fx}, which rules a zero coupling at high energies. By the same token, the coupling increases for decreasing energies, leading to a non-perturbative regime for the strong interactions. Exactly in this low energy regime the bound states, like the nucleons are formed, making it difficult to study, in general. Typically, the description of these bound states in QCD is performed using the lattice approach \cite{Procura:2006bj, BMW:2008jgk, Edwards:2011jj}. Alternative methods for studying QCD bound states include quark models \cite{Capstick:2000qj, Giannini:2015zia}, chiral perturbation theory \cite{Bernard:2003rp}, the Nambu-Jona-Lasinio model \cite{Lawley:2006ps}, and the Schwinger-Dyson equations \cite{Qin:2019hgk}. It is important to mention that the Higgs mechanism \cite{Higgs:1964ia, Higgs:1964pj, Englert:1964et} for the quark masses, although essential, is not enough to explain the nucleon masses. Most of the proton and neutron masses arise from the complex interaction between quarks and gluons (see, {\sl e.g.}, Ref.~\cite{Roberts:2022rxm}). 

String theory was originally proposed as a model for hadronic physics, reproducing Regge behavior, but was unable to describe fixed-angle scattering \cite{Green:1987sp}. This situation changed after the proposal of the AdS/CFT correspondence, which establishes a duality between a gravitational theory in five dimensions and a conformal Yang-Mills theory in four dimensions. Explicitly, a type IIB string theory in five-dimensional anti-de Sitter space times a five-dimensional hypersphere, $\rm AdS_{5}\times S^{5}$, and an extended supersymmetric $\mathcal{N} = 4$ conformal Yang-Mills theory in a four-dimensional flat spacetime \cite{Maldacena:1997re, Witten:1998qj, Gubser:1998bc}. Since then, many models inspired by this gauge/gravity duality breaking conformal invariance and/or supersymmetry have appeared in the literature and are commonly referred to as holographic QCD or AdS/QCD models.

  Nucleon properties, such as their mass spectra, structure functions, and form factors, have been extensively investigated within several holographic frameworks, including hard-wall and soft-wall models \cite{Hong:2006ta, Fang:2016uer, Fang:2016dqm,  Gutsche:2011vb, Gutsche:2017lyu}, deformed AdS models \cite{FolcoCapossoli:2019imm, FolcoCapossoli:2020pks, Contreras:2021epz, Nascimento:2025smf}, the Sakai-Sugimoto model \cite{Sakai:2004cn, Sakai:2005yt, Hata:2007mb, Bayona:2011xj, Ballon-Bayona:2012txi}, and V-QCD approaches \cite{Jarvinen:2022mys, Jarvinen:2022gcc, Deng:2025fpq, Hippelainen:2026izm}. Among these holographic approaches, a particularly relevant class is provided by a model in which the bulk theory is a five-dimensional Einstein-dilaton (ED) gravity. The gravitational part of this model is defined by the Einstein-Hilbert action with an additional scalar field (the dilaton) subjected to a convenient potential and is consistent with Einstein's field equations. This potential is chosen such that in the UV limit, one approaches the boundary of the space, which becomes asymptotically AdS. Furthermore, in the IR limit, this potential  satisfies a general confinement criterion for the dual field theory in flat four dimensions. In particular, this potential can be formulated in such a way that it has a one-to-one correspondence with the QCD beta functions and a non-trivial vacuum \cite{Gursoy:2007cb,Gursoy:2007er}. This model has been used to investigate the meson spectrum, decay constants, and form factors \cite{Chen:2022pgo, Li:2013oda, Ballon-Bayona:2023zal}, as well as the Regge pomeron and odderon trajectories \cite{FolcoCapossoli:2016fzj, Ballon-Bayona:2015wra}. Nevertheless, nucleons within the Einstein-dilaton model have been little explored \cite{Ballon-Bayona:2024yuz, Nascimento:2025ryr}, in which the spectrum and structure functions were investigated.


 The Starobinsky model \cite{Starobinsky1980} is a cornerstone of inflationary cosmology, as it demonstrates that a period of accelerated expansion can arise naturally from geometric corrections to Einstein gravity, without the need for fundamental scalar fields. This gravity model is based on the inclusion of a quadratic curvature term, $\alpha R^2$, in the gravitational action, which can be interpreted as an effective description of quantum corrections at high curvatures. In such a framework, inflation emerges dynamically from the gravitational sector itself, providing a clear link between inflationary dynamics and quantum effects in effective gravity \cite{DeFelice2010, Nojiri2011}. Moreover, this model yields robust predictions for key inflationary observables (such as the scalar spectral index and the tensor-to-scalar ratio) that are in excellent agreement with current cosmological data \cite{Akrami2020}. 
 
 Motivated by this success, a new holographic model has recently been proposed \cite{daSilvaJunior:2025wyn} extending the Einstein-dilaton set up including corrections of order $R^2$ to the usual $R$ term in the Einstein-Hilbert action. This term is the Starobinsky correction in inflationary cosmological models. For this reason, we call this proposal the Starobinsky-dilaton (SD) holographic model, which was used to describe the spectrum of rho vector mesons. Note that these $R^2$ corrections come with a free parameter $\alpha$, such that the bulk space is still asymptotically AdS in the UV, and confining for the dual four-dimensional field theory in the IR. 

In this work, we describe the nucleon spectrum in holographic models with ED and SD gravities. In the ED gravity case, the model has two parameters: $k$ responsible for the conformal symmetry breaking introduced in the scale factor and $\lambda$ introduced in the effective mass in the Schrödinger-like nucleon equations for the Right and Left modes. In the SD gravity model we have three parameters: the same $k$ and $\lambda$ of the ED case together with $\alpha$, which gives the amplitude of the quadratic corrections of the Ricci scalar in the action from the ED case. This allows us to improve the comparison of the nucleon spectra with experimental data and the soft-wall model. 

The outline of this paper is as follows: In Sect.~\ref{ED-SD}, we summarize the ED and SD models and introduce their corresponding parameters used in this work to describe nucleon spectra and wave functions. In Sect.~\ref{einsteinstarobinsky}, we present the ED and SD gravity actions, the field equations, the scale factor, the dilaton fields and the confinement criteria. In Sect.~\ref{nucleonsEDSD}, we introduce the nucleon action and field equations, and obtain a Schrödinger-like equation for their Right and Left modes within both ED and SD frameworks. In Sect.~\ref{modelI}, we discuss the nucleon spectra and wave functions in the ED holographic model with $\lambda$ of the Schrödinger-like effective mass variable. Then, in Sect.~\ref{modelII}, we study the nucleon spectra and wave functions in the SD  holographic model for $\lambda$ fixed and variable. Finally, our conclusions are presented in Sect.~\ref{conclusions}.


\section{Einstein-dilaton and Starobinsky-dilaton models}\label{ED-SD}

In this section, we present a brief introduction to the holographic ED and SD models for nucleons, discussed in this work. The novelties here are mainly two: we introduce a parameter $\lambda$ in the five-dimensional effective mass $M_{\rm eff}$, Eq.~\eqref{URLMeff}, which will be used to improve the fit of nucleon spectra with respect to the experimental data, and the SD models for nucleons. The details of these models are presented in the following sections, as well as the numerical analysis of the nucleon spectra in each case. 

The action, in the Einstein frame, for these models can be written as
\begin{align}
     S =\dfrac{1}{16\pi G_{5}}\int d^{5}x\, \sqrt{-g}\left[f(R)  - \dfrac{4}{3}g^{mn}\partial_{m}\phi\partial_{n}\phi + \ell^{-2}  V(\phi) \right]\,, \label{actiongeral}
\end{align}
where $f(R)$ is some convenient function of the Ricci scalar $R$, which we will take as $f(R)=R+\alpha R^2$, with $\alpha$ being a free parameter of the SD model and, when $\alpha= 0$, we retrieve the traditional ED case. Moreover, $\phi$ denotes the dilaton field, $V(\phi)$ its potential, while $G_{5}$ and $\ell$ represent the five-dimensional Newton constant and the asymptotic AdS radius, respectively. The metric is defined as
\begin{align}
    ds^{2} = \dfrac{1}{\zeta(z, k)^{2}}[dz^{2} + dx_{i}^{2} - dt^{2}]\,;\label{metriczeta}
\qquad\qquad 
    \zeta(z, k) = z\exp\left(\dfrac{2}{3}k\, z^{2}\right)\,. 
\end{align}
where $\zeta(z, k)$ is the scale factor, $i=1,2,3$, and $k$ is the infrared scale, responsible for conformal symmetry breaking. 
Performing the transformations
\begin{align}
    V_{s}(\phi) =V(\phi)\,e^{-{4}\phi/3}\,;\label{transform}
    \qquad\qquad 
    g_{mn}^{s} = g_{mn}\,e^{{4}\phi/3}\,,
\end{align}
with $m,n=0,1,2,3,5$, the above action can be written in the string frame as
\begin{align}
     S =\dfrac{1}{16\pi G_{5}}\int d^{5}x\, \sqrt{-g_{s}}\,e^{-2\,\phi}\left[f(R_{s})  - 4\,g^{mn}\partial_{m}\phi\partial_{n}\phi + \ell^{-2}  V_{s}(\phi) \right]\,,\label{actiongeralSframe}
\end{align}
where $R_s$ is the Ricci scalar evaluated with the metric $g_{mn}^{s}$. 

Introducing fermions in these metrics with the appropriate Dirac action, after decomposing the solutions of the equations of motion into Right and Left modes, one finds a Schrödinger-like equation (see Sections  \ref{einsteinstarobinsky}-\ref{modelI} for more details)
\begin{align}
    \left[- \partial_{z}^{2} +  U_{R/L}\right]\Psi_{R/L} = M_{N^{n}}^{2}\Psi_{R/L}\,,\label{Schrodingerlike}
\end{align}
where $\Psi_{R/L}(z)$ are the eigenfunctions and   %
\begin{align}
  U_{R/L} = \left(\dfrac{M_{\text{eff}}}{\zeta_{s}}\right)^{2} \pm \partial_{z}\left(\dfrac{M_{\text{eff}}}{\zeta_{s}}\right); \qquad \text{with}\qquad  M_{\text{eff}} =\left[\lambda\,\phi^{\prime}\zeta_{s} + m_5\,\zeta_{s}\right]\,, \label{URLMeff}
\end{align}
are the Schrödinger potentials $U_{R/L}$ for the Right and Left modes.  $M_{\text{eff}}$ is the five-dimensional effective mass that depends on the scale factor $\zeta_s$ in the string frame:
\begin{align}
   \zeta_s\equiv  \zeta_{s}(z, k,\alpha) =\zeta(z, k)\,e^{-{2}\phi(z, k,\alpha)/3}\,,\label{zeta_S-frame}
\end{align}
and also on the parameters $m_5$, the usual fermionic mass in five dimensions, and $\lambda$, is an additional parameter which will be conveniently varied to improve the nucleon spectra with respect to experimental data. Note that the dependence of $\zeta_s(z, k,\alpha)$ on $\alpha$, comes through the dilaton profile $\phi=\phi(z, k,\alpha)$ in the SD case, while with $\alpha=0$ in the ED models. By definition, the scale factor $\zeta(z, k)$ appearing in the metric Eq. \eqref{metriczeta} is independent of $\alpha$. 

In the following sections, we are going to discuss the ED and SD models in detail. In sections \ref{modelI} and \ref{modelII} we will study the nucleon spectra and wave functions for some choices of the parameters $k$, $\lambda$, and $\alpha$. We organized these discussions into two models according to the parameters that will be varied or kept fixed: 

\begin{itemize}

    \item [$\bullet$] ED model: Einstein-dilaton with $\lambda$ variable; 

    \item [$\bullet$] SD model A: Starobinsky-dilaton with $\lambda$ fixed. 

     \item [$\bullet$] SD model B: Starobinsky-dilaton with $\lambda$ variable.
    
\end{itemize}

In these models we are going to consider $m_5=5/2,3/2$, so that the conformal dimension $\Delta$ of the nucleon boundary operators in four dimensions will be 9/2 and 7/2, respectively (see Section \ref{nucleonsEDSD} for details).

\section{Einstein-dilaton and Starobinsky-dilaton gravities} \label{einsteinstarobinsky}

In this section, we detail the ED and SD gravity frameworks, presented in brief in the previous section. 

First, we discuss Einstein gravity, formulated in the $\rm AdS_{5}$ spacetime and coupled to a dilaton field, providing an effective holographic description of some non-perturbative properties of QCD. According to the AdS/QCD holographic dictionary, the dilaton is dual to the gauge invariant operator ${\rm Tr}(F^{2})$, which describes the gluon condensate. This setup, commonly called the Einstein-dilaton model, with the appropriate choice of the dilaton profile and its potential, offers a confining framework for dual field theory \cite{Gursoy:2007cb, Gursoy:2007er}.

\subsection{Action and  field equations in ED gravity}\label{einsteindilaton}
The ED action written in the Einstein frame is given by \cite{Ballon-Bayona:2024yuz, Gursoy:2007cb}
\begin{align}
     S =\dfrac{1}{16\pi G_{5}}\int d^{5}x\, \sqrt{-g}\left[R  - \dfrac{4}{3}g^{mn}\partial_{m}\phi\partial_{n}\phi + \ell^{-2}  V(\phi) \right]\,,  \label{edactioneinsteinframe}
\end{align}
where $G_{5}$ the five-dimensional Newton's constant, $R$ is the Ricci scalar, $\phi$ is the dilaton, $V(\phi)$ is the dilaton potential and $\ell$ is the $\rm AdS_{5}$ radius. The above action can also be written in string frame as
\begin{align}
    S =\dfrac{1}{16\pi G_{5}}\int d^{5}x\,\sqrt{-g_{s}}\,e^{-2\,\phi}\left[R_{s} + 4\,g^{mn}\partial_{m}\phi\partial_{n}\phi + \ell^{-2}V_{s}(\phi)\right]\,. \label{edactionstringframe}
\end{align}
This frame, which is analogous to the Jordan frame in general relativity (GR), is related to the Einstein frame via the transformations
\begin{align}
    V_{s}(\phi) &=V(\phi)\,e^{-\frac{4}{3}\phi}\,,\label{potentialstring}\\ 
    g_{mn}^{s} &= g_{mn}\,e^{\frac{4}{3}\phi}\,,\label{metricstring}
\end{align}
with $m,n=0,1,2,3,5$ and $R_s$ is the Ricci scalar defined from the metric $g_{mn}^{s}$. 

Returning to the Einstein frame and varying the action \eqref{edactioneinsteinframe} with respect to the metric tensor and dilaton field, we obtain
\begin{align}
    R_{mn} - \dfrac{R}{2}g_{mn} &= 8\pi G_{5}\, T_{mn}\, ,\label{eqr}\\
     \nabla^{2}\phi + \dfrac{3}{8\ell^{2}}\dfrac{dV}{d\phi} &= 0\,,\label{dilatoneq1}
\end{align}
 with $R_{mn}$ being the Ricci tensor, and $T_{mn}$ the energy-momentum tensor  given by 
\begin{align}
    T_{mn} =\dfrac{8}{3}\partial_{m}\phi\partial_{n}\phi + g_{mn}\mathcal{L}_{\phi}\qquad \text{with}\quad \mathcal{L}_{\phi} = -\dfrac{4}{3}g^{mn}\partial_{m}\phi \partial_{n}\phi + \ell^{-2}\,V(\phi)\,. \label{energymomentumtensor}
\end{align}
In holographic QCD models, the following $5d$ line element is commonly adopted \cite{Ballon-Bayona:2017sxa} 
\begin{align}
    ds^{2} = \dfrac{1}{\zeta(z)^{2}}[dz^{2} + dx_{i}^{2} - dt^{2}]\,,\label{metric5d}
\end{align}
where $i = 1, 2, 3$ denote the spatial coordinates in dual field theory and the scale factor $\zeta(z)$ is written in the Einstein frame (we omit here the constant $\kappa$ introduced in the previous section, which will be discussed in detail in Sect.~\ref{scalefactordilatonfield}). Using this metric, the field equations \eqref{eqr} and \eqref{dilatoneq1} become
\begin{align}
&  12 \left(\dfrac{\zeta^{\prime\, 2}}{\zeta} - \dfrac{\zeta^{\prime\prime}}{\zeta}\right) + 4 \phi^{\prime\,2} - \dfrac{\ell^{-2}}{\zeta^{2}}V = 0\,\label{edeq1}\\
& 12 \,\dfrac{\zeta^{\prime\, 2}}{\zeta^{2}}  - 3\,\dfrac{\zeta^{\prime\prime}}{\zeta} - \dfrac{\ell^{-2}}{\zeta^{2}}V = 0\,,\label{edeq2}\\
& \frac83 \zeta^2 \Big [  \phi'' - 3 \frac{\zeta'}{\zeta} \phi' \Big ] + \ell^{-2} \frac{d V}{d \phi} = 0 \, . \label{dVeq}
\end{align}
where $'=d/dz$ and we considered $8\pi G_{5} = 1$, for simplicity. From Eqs.~\eqref{edeq1} and \eqref{edeq2}, we can eliminate the dependence on the dilaton potential $V$ and get the following equation involving only the scale factor and the dilaton field
\begin{align}
    \zeta^{\prime\prime} - \dfrac{4}{9}\phi^{\prime\, 2}\zeta = 0\,, \label{edpure}
\end{align}
while the dilaton potential $V$ can be obtained from \eqref{edeq2} as
\begin{align}
    V = 12 \zeta^{\prime\,\,2}\,\ell^{2} - 3\zeta^{\prime\prime}\zeta\,\ell^{2}\,.\label{edpotential}
\end{align}
 In the next subsection, we extend the ED framework to the $f(R)$-dilaton case \cite{daSilvaJunior:2025wyn}. This generalization allows us to explore modified gravity scenarios, including higher-order curvature terms in the traditional ED action, under which we derive the field equations of SD gravity.

\subsection{Action and  field equations in SD gravity}\label{starobinskydilaton}

 A simple but powerful modification of GR that introduces an extra scalar degree of freedom is obtained by replacing $R$  in the conventional Einstein-Hilbert action by a general function $f(R)$ \cite{Sotiriou2010, DeFelice2010}. Depending on its specific functional form, this modified gravity model can successfully account for cosmic acceleration and inflation, as well as  strong-field gravitational phenomena, while remaining compatible with a wide range of observational measurements \cite{Capozziello2011, Nojiri2011, Clifton2012, Nojiri2017, NOJIRI2020, ASTASHENOK2015160, ASTASHENOK2020135910, Pretel2022JCAP, Numajiri2023, NOJIRI2025101785}. 
 
 Recently, this idea has been extended to a holographic QCD model based on $f(R)$-dilaton gravity \cite{daSilvaJunior:2025wyn}. In such a formulation, the usual ED framework is generalized by substituting the Ricci scalar in the $5d$ action with an arbitrary function $f(R)$, introducing an additional geometric scalar degree of freedom that works along with the dilaton field $\phi$. This modification provides more  flexibility to reproduce key features of holographic QCD such as the vector meson spectrum. 

To incorporate higher-order curvature corrections into the geometric sector of standard ED holography, the action \eqref{edactioneinsteinframe} can be extended as follows
\begin{equation}
     S =\dfrac{1}{16\pi G_{5}}\int d^{5}x\, \sqrt{-g}\left[f(R)  - \dfrac{4}{3}g^{mn}\partial_{m}\phi\partial_{n}\phi + \ell^{-2}  V(\phi) \right]\,, \label{actioneinsteinfr}
\end{equation}
where $f(R)$ is a generic function of the Ricci scalar $R$. Using the transformations \eqref{potentialstring} and \eqref{metricstring}, the action \eqref{actioneinsteinfr} can be written in the string frame as
\begin{equation}
     S =\dfrac{1}{16\pi G_{5}}\int d^{5}x\, \sqrt{-g_{s}}\,e^{-2\,\phi}\left[f(R_{s})  - 4\,g^{mn}\partial_{m}\phi\partial_{n}\phi + \ell^{-2}  V_{s}(\phi) \right]\,.\label{actionstringfr}
\end{equation}
By varying the action \eqref{actioneinsteinfr} with respect to the metric tensor and dilaton field, we arrive at the following modified field equations
\begin{align}
    f_{R}R_{mn} - \dfrac{1}{2}g_{mn}f + \left(g_{mn}\nabla^{2} - \nabla_{m}\nabla_{n}\right)f_{R} &= 8\pi G_{5}\, T_{mn}\label{eqmotionfr}\,,\\
    \nabla^{2}\phi + \dfrac{3}{8\ell^{2}}\dfrac{dV}{d\phi} &= 0\,,\label{dilatoneq}
\end{align}
where $f_{R} = df(R)/dR$. Using the line element \eqref{metric5d}, the field equations \eqref{eqmotionfr} become
 \begin{align}
   R^{\prime\prime}f_{RR} + R^{\prime\,\, 2}f_{RRR} - 6\dfrac{\zeta^{\prime}}{\zeta} R^{\prime}f_{RR} - \dfrac{1}{\zeta^{2}}(f + \ell^{-2}V - R f_{R}) +  3\left(4\dfrac{\zeta^{\prime\,2}}{\zeta^{2}} - \dfrac{\zeta^{\prime\prime}}{\zeta}\right)f_{R} = 0\,, \label{fReqmotion1}\\
   4\left(R^{\prime\, \prime}f_{RR} + R^{\prime\, 2}f_{RRR} + \phi^{\prime\, 2}\right) + 12 \left(\dfrac{\zeta^{\prime\, 2}}{\zeta^{2}} - \dfrac{\zeta^{\prime\prime}}{\zeta}\right)f_{R} - \dfrac{1}{\zeta^{2}}(f + \ell^{-2}V - Rf_{R}) = 0\,, \label{fReqmotion2}\\
   4\left(R^{\prime\, \prime}f_{RR} + R^{\prime\, 2}f_{RRR} - 3 \dfrac{\zeta^{\prime}}{\zeta}R^{\prime}f_{RR}\right) + \dfrac{1}{\zeta^{2}}R f_{R} - \dfrac{5}{2\zeta^{2}}(f + \ell^{-2}V) + 2\phi^{\prime\, 2} = 0\,. \label{fReqmotion3}
 \end{align}

In particular, we are interested in the well-known Starobinsky model, characterized by a quadratic form
\begin{align}
    f(R) = R + \alpha R^{2}\,,\label{starobinsky}
\end{align}
where $\alpha$ is a free parameter of the model. Proposed by Alexei Starobinsky in 1980 \cite{Starobinsky1980}, it shows that gravity itself can drive a phase of cosmic inflation through the dynamics of an emergent scalar degree of freedom often called the scalaron (which emerges dynamically because the higher-curvature term makes the theory equivalent to a scalar–tensor theory in the Einstein frame). This mechanism produces a smooth and prolonged period of accelerated expansion and naturally generates primordial density perturbations consistent with observations of the cosmic microwave background (CMB). Its predictions agree strikingly well with modern data from missions such as {\it Planck} CMB anisotropy measurements \cite{Akrami2020}, making it one of the simplest and most empirically successful inflationary scenarios. Motivated by these advances, our subsequent analyzes are based on the quadratic function \eqref{starobinsky}.

In this case, one finds that Eqs.~\eqref{fReqmotion1}-\eqref{fReqmotion3}, reduce to 
\begin{align}
     3\left(4\dfrac{\zeta^{\prime\,2}}{\zeta^{2}} - \dfrac{\zeta^{\prime\prime}}{\zeta}\right)\left(1 + 2\alpha R\right) + 2\alpha(R^{\prime\, \prime} - 6\dfrac{\zeta^{\prime}}{\zeta} R^{\prime}) =\dfrac{1}{\zeta^{2}}(\ell^{-2}V - \alpha R^{2})\,, \label{eqstarobinsky1}\\
     12 \left(\dfrac{\zeta^{\prime\, 2}}{\zeta^{2}} - \dfrac{\zeta^{\prime\prime}}{\zeta}\right)(1 + 2\alpha R) +  4(2\alpha R^{\prime\prime} + \phi^{\prime\,\, 2}) = \dfrac{1}{\zeta^{2}}(\ell^{-2}V - \alpha R^{2})\,, \label{eqstarobinsky2}\\
     16\alpha\left(R^{\prime\prime} -  3\dfrac{\zeta^{\prime}}{\zeta} R^{\prime}\right) + 2R(1 + 2\alpha R)\dfrac{1}{\zeta^{2}} -  5(R + \alpha R^{2} + \ell^{-2}V)\dfrac{1}{\zeta^{2}} + 4\phi^{\prime\,\, 2} = 0\,,\label{eqstarobinsky3} 
 \end{align}
where the Ricci scalar is given by
\begin{align}
    R = 8\,\zeta^{\prime\prime}\zeta - 20\,\zeta^{\prime\,2}\,.\label{ricciscalar}
\end{align}
 From Eqs.~\eqref{eqstarobinsky1} and \eqref{eqstarobinsky2}, one obtains 
\begin{align}
  9(1 + 2\alpha R) \dfrac{\zeta^{\prime\prime}}{\zeta}  - 6\alpha \left(2\,\dfrac{\zeta^{\prime}}{\zeta}\,R^{\prime} + R^{\prime\prime}\right)  - 4\phi^{\prime\, 2} =0\,,\label{eqstarobinsky4}
\end{align}
and substituting the Ricci scalar, Eq.~\eqref{ricciscalar}, into this equation, one finds 
\begin{align}
     4\phi^{\prime\, 2}\zeta = 24\alpha \zeta^{\prime\, 2} \zeta^{\prime\prime} + 48\alpha \zeta^{\prime\prime\prime}\zeta^{\prime}\zeta - 48\alpha\zeta^{\prime\prime\prime\prime}\zeta^{2} + 336\alpha \zeta^{\prime\prime\, 2}\zeta + 9\zeta^{\prime\prime}\,.\label{eqstarobinsky5}
\end{align}
This is a generalized version of Eq.~\eqref{edpure} obtained in the ED case. The corresponding Starobinsky potential can be determined from 
Eq.~\eqref{eqstarobinsky1} employing the Ricci scalar \eqref{ricciscalar}, namely 
\begin{align}
    V =& \bigg[\left(12 \zeta^{\prime\,\,2} - 3\zeta^{\prime\prime}\zeta\right)\left(1 + 2\alpha\left(8\zeta^{\prime\prime}\zeta - 20\zeta^{\prime\,2}\right)\right) 
    \nonumber \\
    &+ 2\alpha\left(8\zeta^{\prime\,\prime\,\prime\,\prime}\zeta^{\prime\,\prime\,\prime} - 32\zeta^{\prime\,\prime\, 2}\zeta^{2} - 24\zeta^{\prime\,\prime\,\prime}\zeta^{\prime}\zeta^{2}\right) \nonumber \\
    &- 12\alpha\zeta^{\prime}\zeta\left(8 \zeta^{\prime\,\prime\,\prime}\zeta - 32\zeta^{\prime}\zeta^{\prime\,\prime}\right)\bigg] + \alpha\left(8\zeta^{\prime\prime}\zeta - 20\zeta^{\prime\,2}\right)^{2}\,, \label{Starobinskypotential}
\end{align}
where, for simplicity, we considered $\ell = 1$. Note that the results of this section reduce to the previous one taking the limit $\alpha\to0$, which means $f(R) \to R$. In particular, the above potential reproduces Eq.~\eqref{edpotential} in this limit.

In the next subsection, starting with a given scale factor, we obtain explicitly the profiles of the dilaton fields and the corresponding potentials in Einstein- and Starobinsky-dilaton models.


\subsection{Scale factor and dilaton fields in Einstein- and Starobinsky-dilaton models}\label{scalefactordilatonfield}

In this subsection, we introduce the scale factor and the dilaton fields responsible for deforming the $\rm AdS_{5}$ geometry and, consequently, breaking the conformal symmetry. The same scale factor is adopted for both the ED and SD models, while the corresponding dilaton fields differ due to the $\alpha$-dependent corrections inherent to the Starobinsky framework. 
 
In order to solve the field equations in ED and SD gravities of the previous sections, we consider the following profile for the scale factor
\begin{align}
     \zeta(z, k) = z\exp\left(\dfrac{2}{3}k\, z^{2}\right)\,. \label{zeta}
\end{align}
where $k$ is the parameter associated with the mass scale. This scale factor has been used in ED models to study the spectrum of glueballs, vector mesons, and nucleons \cite{Gursoy:2007er, Ballon-Bayona:2024yuz}, and more recently in the $f(R)$-dilaton framework for the analysis of the vector meson spectrum \cite{daSilvaJunior:2025wyn}.
 
 By substituting this scale factor into the ED equation \eqref{edpure}, we obtain the exact solution for the dilaton field 
\begin{align}
    \phi(z, k) = \dfrac{1}{2}\sqrt{k}\,z\,\sqrt{9 + 4\,k\,z} + \dfrac{9}{4}\sinh^{-1}\left(\dfrac{2}{3}\sqrt{k}z\right)\,.\label{dilatonfielded}
\end{align}
The scale factor, Eq.~\eqref{zeta}, and the above dilaton field admit the following expansion near the boundary ($z\to 0$)
\begin{align}
    \zeta(z, k) &= z + \dfrac{2\, k\, z^{3}}{3} + \dfrac{2\,k^{2} z^{5}}{9} + \cdots;\\
    \phi(z, k) 
    &= 3\,k^{1/2}\left(z
     + \dfrac{2\,kz^{3}}{27} - \dfrac{2\,k^{2}z^{5}}{405} + \cdots\right).
    \label{dilatonexpansionalt2}
\end{align}

Analogously, in the SD case, substituting the scale factor \eqref{zeta} into the field equation \eqref{eqstarobinsky5}, we derive the dilaton field incorporating the $\alpha$-dependent corrections, namely
\begin{align}
     \phi(z,k,\alpha) = \dfrac{1}{3}\int_{0}^{z}\, d\bar{z} & \left[\Big(10624 k^{3} \bar{z}^{4}\alpha + 2560 k^{4} \bar{z}^{6}\alpha\Big)
     \exp\left(\frac{4}{3}k\,\bar{z}^{2}\right) 
     \right. \nonumber  \\
     & \left. + 81 k\left(1 + 8 \alpha\exp\left(\frac{4}{3}k\,\bar{z}^{2}\right)\right) \right.  \nonumber  \\
    &\left. + 12 k^{2} \bar{z}^{2}\left(3 + 968 \alpha \exp\left(\frac{4}{3}k\,\bar{z}^{2}\right)\right)\right]^{1/2}\,.\label{dilatonalpha}
\end{align}
If one sets  $\alpha = 0$ in the above equation, one recovers  Eq.~\eqref{dilatonfielded} for the ED case in closed form. The expansion near the boundary of this SD field is given by 
\begin{align}
    \phi(z, k,\alpha) =  \sqrt{k(1 + 8\alpha)} \left( 3z + \dfrac{2\,k\, z^{3} (3 + 1040\,\alpha)}{3(9 + 72\alpha)} + \cdots\right).\label{dilatonalphaexpansion}
\end{align}

\begin{figure}[htp!]%
    \centering
    {{\includegraphics[width=7.2cm]{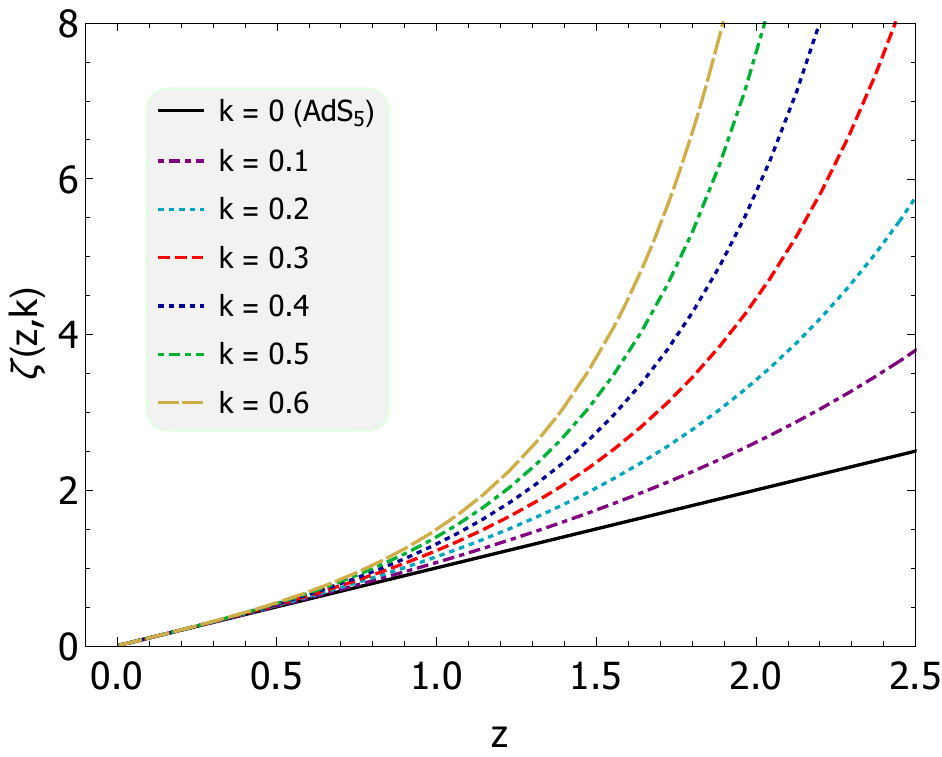}}}%
    \caption{Scale factor $\zeta(z, k)$ as function of holographic coordinate $z$ for some values of $k$.}%
    \label{Plot:zetav3}%
\end{figure}

The behavior of the scale factor $\zeta(z,\kappa)$ for some values of $k$ is illustrated in Fig.~\ref{Plot:zetav3}. Indeed, for $k = 0$, the $\rm AdS_{5}$ solution is recovered and is represented by the solid black curve. The remaining curves with $k = 0.1$ to $k = 0.6$ correspond to deformations of the $\rm AdS_{5}$ geometry, which imply a breaking of the conformal symmetry in the dual field theory. For instance, in subsection \ref{einsteindilatonlambdavarying}, we are going to calculate the nucleon spectrum in ED gravity, for $k$ values ranging from $0.3$ to $0.6$. This variation results from changes in the $\lambda$ parameter.

\begin{figure}[htp!]%
    \centering
    {{\includegraphics[width=7.2cm]{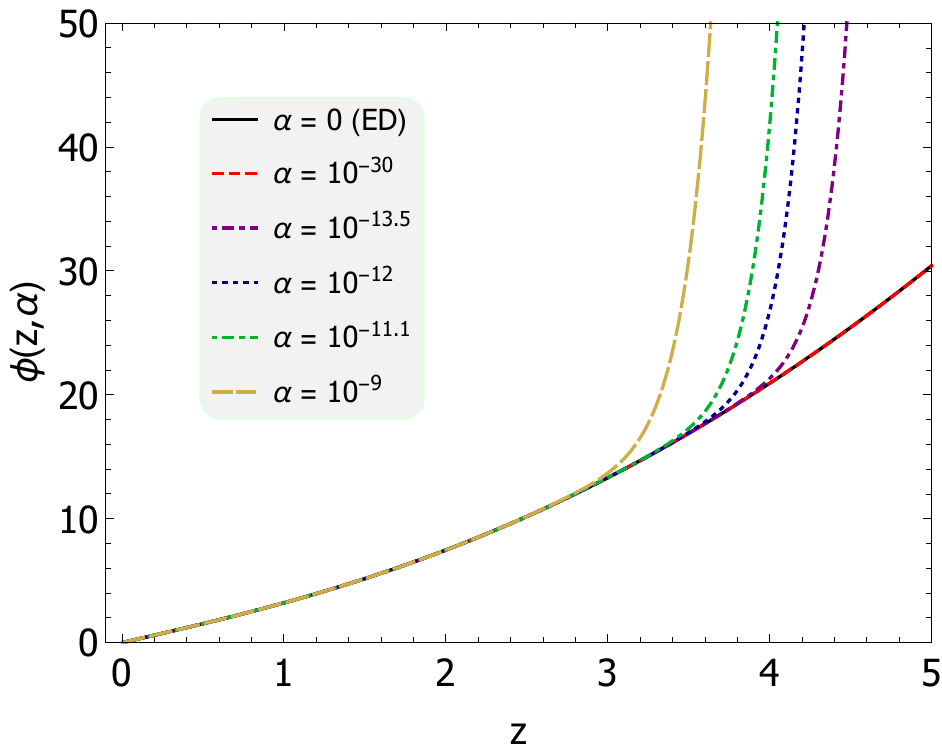}}
    \hskip .5cm {\includegraphics[width=7.2cm]{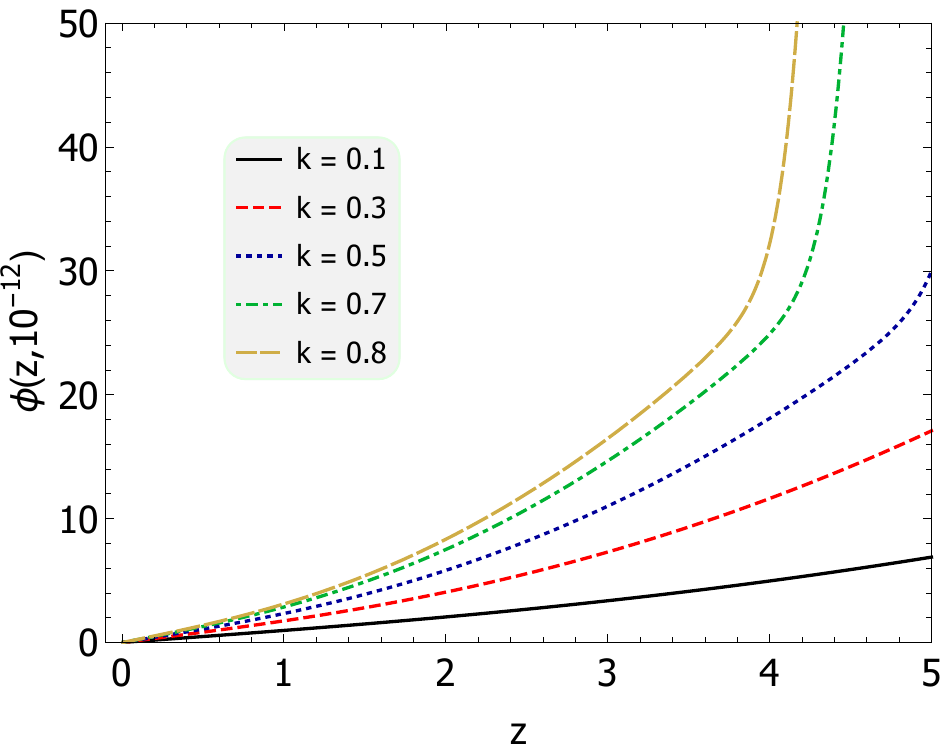}}}%
    
     \caption{Dilaton field profiles $ \phi(z,\alpha)$ in SD framework as a function of the holographic coordinate $z$. {\bf Left panel}: several values of $\alpha$ are shown with $k =1$ fixed.
 {\bf Right panel}: several $k$ values at fixed $\alpha = 10^{-12}$.}%
    \label{Plot:Starobinskydilatonv2}%
\end{figure}

 The behavior of the dilaton field in the SD gravity in two representative cases is displayed in Fig.~\ref{Plot:Starobinskydilatonv2}. The left panel corresponds to a fixed value of $k = 1$ with varying $\alpha$, while the right panel shows the results for fixed $\alpha = 10^{-12}$ with varying $k$. In the left panel of this figure, one sees that for $\alpha = 0$, corresponding to the ED case, the dilaton grows smoothly with the holographic coordinate $z$. As $\alpha$ increases, the profiles progressively depart from the ED solution, with the dilaton growing more rapidly at higher values of $z$, indicating that higher values of $\alpha$ enhance the deformation of the background geometry. In the right panel of this figure, one observes that increasing $k$ accelerates the growth of the dilaton with $z$, showing that $k$ amplifies the curvature effects and strengthens the dilaton field.  Their combined influence determines the overall profile of the dilaton field and the degree of deviation from the pure $\rm AdS_{5}$ background.

The dilaton potential for the ED case is obtained by substituting Eq.~\eqref{zeta} into Eq.~\eqref{edpotential}. Near the boundary, the expansion of this potential takes the form of

 \begin{align}
     V(z, k) = e^{\frac{4}{3}k\, z^{2}} \left( 12\,  + 20\, k\,z^{2}\,  + 16\,k^{2} z^{4}\,   + \cdots \right). 
     \label{edpotentialexpansion}
 \end{align}
 
\begin{figure}[H]%
    \centering
    {{\includegraphics[width=7.24cm]{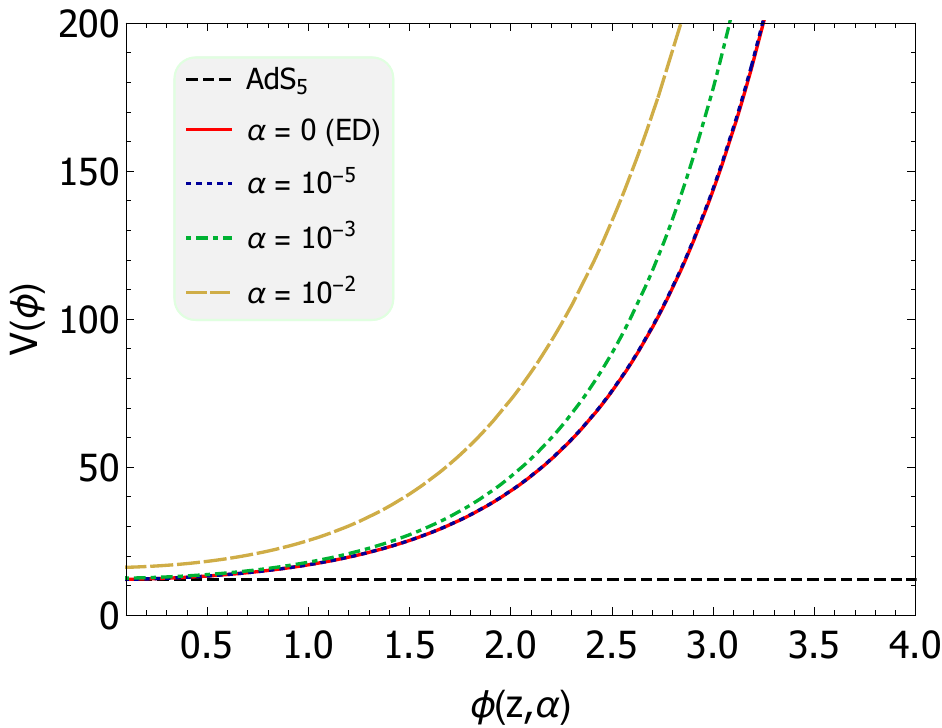}}
    \hskip 0.5cm 
    {\includegraphics[width=7.2cm]{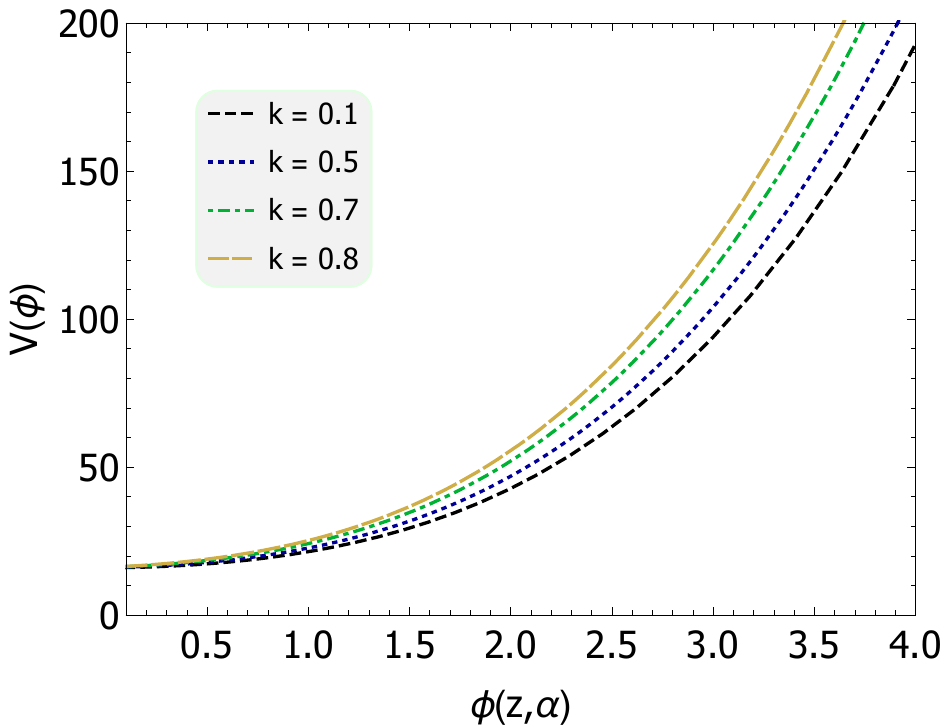}}}%
    
    \caption{SD potential $V(\phi)$ $=V(z, k, \alpha)$, Eq.~\eqref{Starobinspotentialexpansion}, as a function of the dilaton field $\phi(z,\alpha)$. {\bf Left panel}: several $\alpha$ values at fixed $k = 1$ and $\alpha= k =0$ (AdS case, black dashed line).
 {\bf Right panel}: fixed $\alpha = 10^{-2}$ and various values of the parameter $k$.}%
    \label{Plot:Starobinskydilatonpotential}%
\end{figure}

For $k = 0$, the potential reduces to a constant, which is related to the cosmological constant of the AdS space. The SD potential can be obtained by substituting Eq.~\eqref{zeta} into Eq.~\eqref{Starobinskypotential}. Near the boundary, the potential admits the following expansion
\begin{align}
    V(z, k, \alpha) = 12 - 80\alpha + \dfrac{5120}{3}\,k^{3}\alpha z + 36\,k\,z^{2}(1 + 8\,\alpha) + \cdots\label{Starobinspotentialexpansion}
\end{align}
One can observe that this expression reduces to the previous one when  $\alpha=0$. Further, 
the $\alpha$-dependent corrections of this potential first appear in its  second term, which does not involve the parameter $k$. In contrast, in the subsequent terms both parameters appear. Figure~\ref{Plot:Starobinskydilatonpotential} shows this  potential as a function of the dilaton field for fixed $k = 1$ and several values of $\alpha$, including the $\rm AdS_{5}$ solution corresponding to $\alpha= k = 0$, represented by the black dashed line (left panel), and for fixed $\alpha = 10^{-2}$ with varying $k$ (right panel). Note that for $\alpha = 10^{-5}$, the SD solution practically coincides with the ED one, while for $\alpha = 10^{-3}$ and $\alpha = 10^{-2}$, the potential increases significantly compared to the ED case, indicating that Starobinsky gravity enhances the potential. On the other hand, for fixed $\alpha = 10^{-2}$ and varying $k$, the potential also becomes stronger. Both parameters $\alpha$ and $k$ break the conformal invariance of dual field theory independently, although there are crossing terms $\alpha\,k$ in the potential.  

The following subsection examines the impact of the parameter $k$ on the confinement criterion on ED model and parameters $\alpha$ and $k$ on SD models.


\subsection{Confinement}\label{confinement}

Let us start this discussion with the ED model, proposed in Ref.~\cite{Gursoy:2007er}, which establishes a  confinement criterion for this model based on the possible behaviors of QCD beta functions, the superpotential and the metric compatible with the Wilson loop. Here, we concentrate on the metric-based confinement criterion and analyze the impact of the parameters $\alpha$ and $k$ on it. The Wilson loop is a gauge-invariant non-local operator, defined as the trace of the gauge holonomy around a closed contour. This operator characterizes the energy of a static heavy quark-antiquark pair. In the context of the AdS/CFT correspondence, the rectangular Wilson Loop associated with the $\mathcal{N} = 4$ super Yang-Mills theory can be written as \cite{Maldacena:1998im, Brandhuber:1998er}
\begin{align}
    \langle W (\mathcal{C})\rangle = e^{ - S_{\text{NG}}(\mathcal{C})}\,,
\end{align}
where $\langle W (\mathcal{C})\rangle$ is the VEV of the Wilson loop operador as a function of the closed curve $\mathcal{C}$ and $S_{\text{NG}}(\mathcal{C})$ is the on-shell Nambu-Goto action, which describes the area of the world sheet. The Nambu-Gotto action is related to the energy of the pair in ED gravity, which can be written as
\begin{align}
    E(d) = T\,\zeta_{s}^{- 2}(z^{\star}, k)\, d\,, \label{quarkantiquark}
\end{align}
with $E(d)$ being the energy of the pair as a function of the distance of separation $d$, $T$ is the tension of the string and $\zeta_{s}(z^{\star}, k)$ is the scale factor written in the string frame\footnote{The energy of the quark-antiquark pair in ED gravity is frequently written in terms of warp factor, $A_{s}(z)$, in string frame, where $\zeta_{s}^{-2}(z) = e^{2A_{s}(z)}$; see Ref.~\cite{Ballon-Bayona:2017sxa}.}. At $z = z^{\star}$, the scale factor is at its minimum value. Once this minimum value of the scale factor is non-zero, the model is confining \cite{Kinar:1998vq}. The relation between the scale factor in the string-frame and in the Einstein frame can be written as
\begin{align}
    \zeta_{s}(z, k) =\zeta(z, k)\,e^{-\frac{2}{3}\phi(z, k)}\,.\label{scaleframesed}
\end{align}

\begin{figure}[H]%
    \centering
    {{\includegraphics[width=7.2cm]{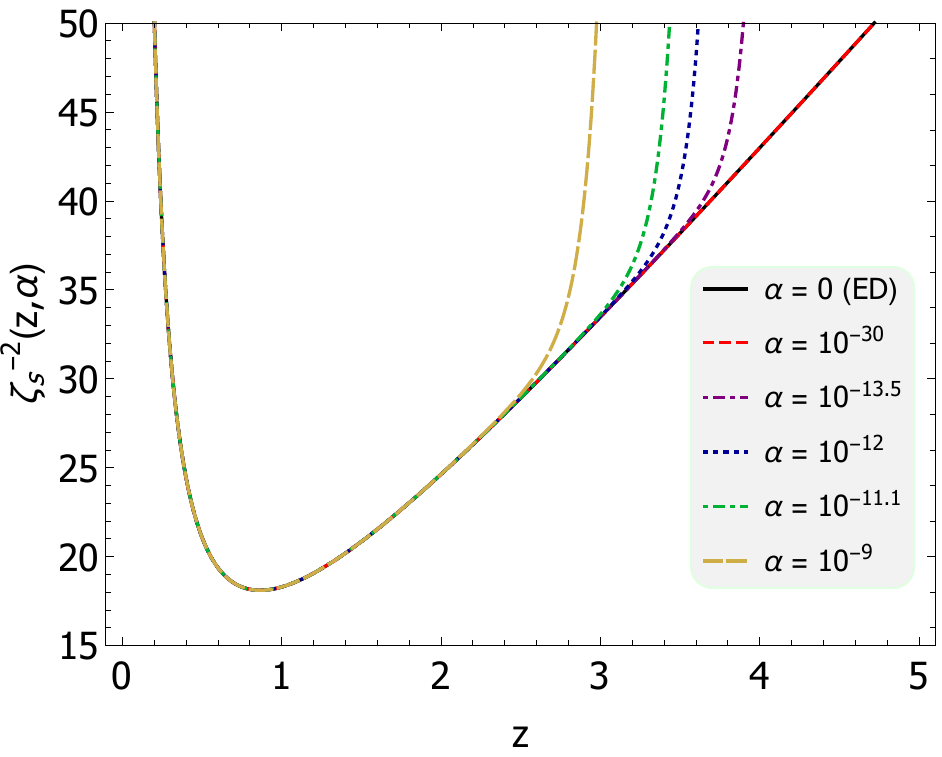}}
    \hskip 0.5cm 
    {\includegraphics[width=7.2cm]{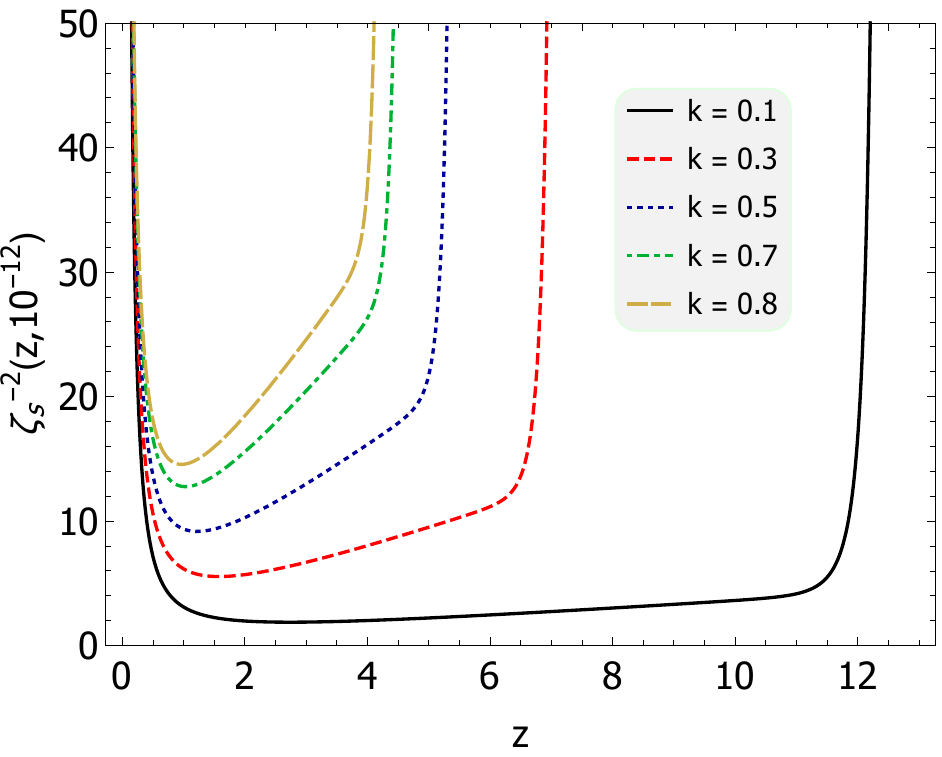}}}%
    \caption{String-frame scale factor $\zeta_{s}^{-2}(z, k,\alpha)$ versus the holographic coordinate $z$ for different $\alpha$ and $k$ values. {\bf Left panel}: fixed $k = 1$.
 {\bf Right panel}:  fixed $\alpha = 10^{-12}$.}%
    \label{Plot:Scalefactor}%
\end{figure}

Next, we move to SD gravity, where the energy of the quark-antiquark pair \eqref{quarkantiquark}, becomes
\begin{align}
    E(d) = T\,\zeta_{s}^{- 2}(z^{\star},k, \alpha)\,d\label{quarkantiquarkalpha}
\end{align}
where $\zeta_{s}^{- 2}(z^{\star},k, \alpha)$ in this case depends on the parameter $\alpha$, since the transformation \eqref{scaleframesed} becomes
\begin{align}
\zeta_{s}(z, k, \alpha) =\zeta(z, k)\,e^{-\frac{2}{3}\phi(z, k, \alpha)}\,\label{scaleframessd}.  
\end{align}
The behavior of the scale factor $\zeta_{s}^{-2}(z,k,\alpha)$ in the string frame is shown in Fig.~\ref{Plot:Scalefactor}, for two configurations: the left panel corresponds to $k = 1$ fixed with varying $\alpha$, and the right plot displays $\alpha = 10^{-3}$ fixed with varying $k$. 

According to the confinement criteria for field theories dual to general gravitational theories with diagonal metrics \cite{Kinar:1998vq}, the scale factor in the string frame must exhibit a non-zero minimum. In the left panel of Fig.~\ref{Plot:Scalefactor}, we reproduce this behavior for ED ($\alpha = 0$) and SD gravities. The purple, blue, green, and yellow curves corresponding to $10^{-13.5}$, $10^{-12}$, $10^{-11.1}$, and $10^{- 9}$, respectively, also display a minimum, ensuring confinement for SD gravity. We further investigated negative values of $\alpha$ and found that the resulting corrections do not lead to confinement; for this reason, we restrict our analysis to positive values of $\alpha$. On the other hand, from a cosmological standpoint, the parameter $\alpha$ must be positive to ensure theoretical consistency and stability \cite{Sotiriou2010, DeFelice2010}. A negative $\alpha$ makes the scalaron tachyonic and can violate the ghost-free condition $f_{R}>0$ in positive-curvature regimes, leading to instabilities and an ill-defined kinetic sector. From a dynamical perspective, only $\alpha>0$ yields a stable inflationary potential capable of supporting a sustained slow-roll phase and generating primordial perturbations consistent with CMB observations. Consequently, both theoretical and observational considerations require $\alpha$ to be positive in viable Starobinsky cosmology.

In the right plot of Fig.~\ref{Plot:Scalefactor}, we show the behavior of the scale factor as a function of the radial holographic coordinate for several values of $k$ with $\alpha = 10^{- 12}$ held fixed. This value of $\alpha$ is chosen to appreciate the onset of nontrivial SD effects, as indicated in the left panel of the same figure. The conformal symmetry is broken for any $k\not=0$. It is also observed that, as $k$ increases from $0.1$ to $0.8$, the breaking of the conformal symmetry becomes progressively stronger, and the curves develop a more pronounced minimum for higher values of $k$. 

Near the boundary, the string-frame scale factor appearing in Eq.~\eqref{quarkantiquark} for ED gravity admits the following expansion:
\begin{align}
    \zeta_{s}^{-2}(z, k)  = \dfrac{1}{z^{2}} + \dfrac{4\sqrt{k}}{z} + \dfrac{20\,k}{3} + \cdots\label{zetastringed}
\end{align}
and for SD gravity, the scale factor in \eqref{quarkantiquarkalpha} near the boundary becomes
\begin{align}
    \zeta_{s}^{-2}(z, k, \alpha)  = \dfrac{1}{z^{2}} + \dfrac{4\sqrt{k(1 + 8 \alpha)}}{z} + \dfrac{4}{3}k(5 + 48\alpha) + \cdots\,,\label{zetastringsd}
\end{align}
such that for $\alpha = 0$ in \eqref{zetastringsd} naturally leads to \eqref{zetastringed}. The $\alpha$-dependent contributions, which begin to emerge from the second term in Eq.~\eqref{zetastringsd}, are responsible for the corrections shown in the left panel of Fig.~\ref{Plot:Scalefactor}, which maintain the confinement criterion established in Ref.~\cite{Gursoy:2007er}. 

Once the ED and SD gravity frameworks have been established, we proceed in the next section to investigate the nucleon sector within these two settings.


\section{Nucleons in Einstein- and Starobinsky-dilaton models}
\label{nucleonsEDSD}

In this section, we start the discussion of the nucleon spectra in the ED and SD models. We are interested in describing the following nucleon states: $N(938)$, $N(1440)$, $N(1710)$, and $N(1880)$. All of these states have positive parity. We start by presenting the Dirac action, field equations, and the associated Schrödinger equation. The nucleon spectrum is computed in two distinct scenarios: ED model: $\lambda$ variable (to be defined in Eq.~\eqref{potentialed}) in ED gravity; and SD model: $\lambda$ fixed and variable in Starobinsky-dilaton gravity (SDA and SDB, respectively). In each model, our results are compared with available experimental data and soft-wall model.

To describe nucleons, let us start with the Spin-$\frac{1}{2}$ field action \cite{Ballon-Bayona:2024yuz} written in  the string frame
\begin{align}
    I = G_{F}\int d^{5}x\, \sqrt{-g_{s}}\,\bar{\chi}\left(\dfrac{i}{2}\slashed{D} + c.c - i M_{\text{eff}}\right)\chi +  G_{F}\int d^{4}x\,\sqrt{-h_{s}}\,\bar{\chi}\,\chi\,,\label{diracaction}
\end{align}
where the first term describes the Dirac field $\chi$ as well as its conjugate $\bar{\chi}$, in the bulk of ${\rm AdS}_{5}$ spacetime, $M_{\text{eff}}$ denotes the effective mass, which receives contributions from derivatives of both the dilaton field and the warp factor and $g_{s}$ is the metric, while the second is the boundary term \cite{Henneaux:1998ch}, required to ensure a well-defined variational principle and plays an important role in the derivation of correlation functions and $h_{s}$ is the induced metric. In the action \eqref{diracaction}, the coupling constant $G_{F} = 2/\pi^{4}$ was fixed in \cite{Ballon-Bayona:2024yuz} to reproduce the two-point function of perturbative QCD \cite{Cohen:1994wm}. 

By varying the action \eqref{diracaction} with respect to $\bar{\chi}$ and $\chi$, one obtains
\begin{align}
    \left(\dfrac{i}{2}\slashed{D} + c.c - i M_{\text{eff}}\right)\chi = 0\,,\label{chi}\\
     \bar{\chi}\left(\dfrac{i}{2}\overleftarrow{\slashed{D}} + c.c + i M_{\text{eff}}\right) = 0\,,\label{barchi}
\end{align}
with the operator $\slashed{D}$ given by
\begin{align}
    \slashed{D} =\Gamma^{N}D_{N} = e^{N}_{A}\Gamma^{A}\left(\partial_{N} + \dfrac{1}{4}\omega_{N}^{AB}\Gamma_{AB}\right)\,,\label{covariantderivative} 
\end{align}
where $N$ is the index of the $5d$ curved spacetime and the indices $A, B$ to the $5d$ flat spacetime. The operador \eqref{covariantderivative} contains the veilbein $e_{A}^{N}$ and the spin connection $\omega_{N}^{AB}$. In this work, we will consider the spin connection  metric-compatible and torsion-free. The veilbein and spin connection written in terms of scale factor are given by
\begin{align}
    e_{A}^{N} &=\delta_{A}^{N}\zeta_{s}(z)\,,\label{veilbein}\\
    \omega_{N}^{AB} &= \dfrac{\zeta_{s}^{\prime}(z)}{\zeta_{s}(z)}\left(\delta_{z}^{B}\delta_{N}^{A} - \delta_{z}^{A}\delta_{N}^{B}\right)\,,\label{spinconnection}
\end{align}
where the scale factor in string frame ${\zeta_{s}(z)}$ in this section could be either given by Eq.~\eqref{scaleframesed} in ED model, or by Eq.~\eqref{scaleframessd} in SD gravity. 

Using Eqs.~\eqref{veilbein} and \eqref{spinconnection} into \eqref{covariantderivative}, one can rewrite Eq.~\eqref{chi} as
\begin{align}
    \left(\zeta_{s}(z)\,\Gamma^{A}\partial_{A} - 2\zeta_{s}^{\prime}(z)\gamma^{z} + M_{\text{eff}}\right)\chi = 0\,.\label{chiexpand}
\end{align}
The Dirac spinor $\chi$ can be written in terms of Right and Left components as
\begin{align}
    \chi(x,z) = \chi_{R}(x,z) + \chi_{L}(x,z)\,,\label{rightleft}
\end{align}
which, in turn, satisfy the following property:
\begin{align}
    \gamma^{z}\chi_{R/L} = \pm \chi_{R/L}\,.\label{chiralproperty}
\end{align}
Returning with these results to Eq.~\eqref{chi}, we obtain the following coupled equations
\begin{align}
    \zeta_{s}\gamma^{\mu}\partial_{\mu}\chi_{R} &= \left(\zeta_{s}\partial_{z} - 2\zeta_{s}^{\prime} + M_{\text{eff}}\right)\chi_{L}\,,\label{coupledrighileft}\\
    \zeta_{s}\gamma^{\mu}\partial_{\mu}\chi_{L} &= -\left(\zeta_{s}\partial_{z} - 2\zeta_{s}^{\prime} + M_{\text{eff}}\right)\chi_{R}\,,\label{coupledleftright}
\end{align}
where we decomposed  the gamma matrix  $\Gamma^{A} = (\gamma^{\mu}, \gamma^{z})$ and the derivatives as $\partial_{A} = (\partial_{\mu}, \partial_{z})$. Applying the operator 
$\gamma^{\mu}\partial_{\mu}$ on the right of Eq.~\eqref{coupledrighileft} and using \eqref{coupledleftright}, we get the decoupled equations
\begin{align}
    \Box \chi_{R/L}(x, z) = -\left(\partial_{z} - 2\dfrac{\zeta_{s}^{\prime}}{\zeta_{s}} \pm \dfrac{M_{\text{eff}}}{\zeta_{s}}\right)\left(\partial_{z} - 2\dfrac{\zeta_{s}^{\prime}}{\zeta_{s}} \mp \dfrac{M_{\text{eff}}}{\zeta_{s}}\right)\chi_{R/L}(x, z)\,.\label{decoupledchi}
\end{align}
The Left and Right Dirac spinors admit a Kaluza-Klein expansion \cite{Gutsche:2011vb} of the form
\begin{align}
    \chi_{R/L}(x,z) =\sum_{n = 0}^{\infty} f_{R/L}^{n}(z)\hat{\chi}_{R/L}(x)\,,\label{kkexpansion}
\end{align}
where $f_{R/L}^{n}(z)$ are the Kaluza-Klein (KK) modes and $\hat{\chi}_{R/L}(x)$ are $4d$ Dirac fields. Substituting this result into the decopupled equations \eqref{decoupledchi}, we get the following for the KK modes 
\begin{align}
    \left[\partial_{z}^{2} - 4\dfrac{\zeta_{s}^{\prime}}{\zeta_{s}}\partial_{z} + 2\left(\dfrac{\zeta_{s}^{\prime\, 2}}{\zeta_{s}^{2}} - \dfrac{\zeta_{s}^{\prime\prime}}{\zeta_{s}}\right) + 4 \dfrac{\zeta_{s}^{\prime\,2}}{\zeta_{s}^{2}} \mp \partial_{z}\left(\dfrac{M_{\text{eff}}}{\zeta_{s}}\right) - \dfrac{M^{2}_{\text{eff}}}{\zeta_{s}^{2}} + M_{N^{n}}\right]f_{R/L}^{n}(z) = 0\,,\label{decoupledmodes}
\end{align}
where $M_{N^{n}}$ denotes the mass of nucleons. These equations are valid for both ED and SD gravities, depending on the choice of the scale factor in the string frame, $\zeta_{s}(z, k)$ or  $\zeta_{s}(z, k, \alpha)$, 
Eqs.~\eqref{scaleframesed} and \eqref{scaleframessd}, respectively. 

Rewriting the KK modes 
$f_{R/L}^n(z)$, Eq.~\eqref{kkexpansion}, as
\begin{align}
    f_{R/L}^{n}(z) =\zeta_{s}^{2}(z,\kappa)\Psi_{R/L}(z)\,,\label{bogoliubov}
\end{align}
and substituting this relation into Eq.~\eqref{decoupledmodes}, one finds a Schrödinger-like equation 
\begin{align}
    \left[- \partial_{z}^{2} +  U_{R/L}\right]\Psi_{R/L} = M_{N^{n}}^{2}\Psi_{R/L}\,,\label{nucleonschrodingereqed}
\end{align}
where $\Psi_{R/L}(z)$ are the eigenfunctions and   %
\begin{align}
  U_{R/L} = \left(\dfrac{M_{\text{eff}}}{\zeta_{s}}\right)^{2} \pm \partial_{z}\left(\dfrac{M_{\text{eff}}}{\zeta_{s}}\right); \qquad \text{with}\qquad  M_{\text{eff}} =\left[\lambda\,\phi^{\prime}\zeta_{s} + m_5\,\zeta_{s}\right]\,,\label{potentialed}
\end{align}
are the Schrödinger potentials for the Right and Left modes, $U_{R/L}$, and the effective mass $M_{\text{eff}}$. This effective mass depends on the scale factor $\zeta_s$ and also on the parameters  $m_{5}$, the usual fermionic mass in five dimensions, and $\lambda$, which will be conveniently varied as an additional parameter in the ED and SD models.

According to the detailed discussion in Ref.~\cite{Ballon-Bayona:2024yuz}, an effective mass of the form given in Eq.~\eqref{potentialed} is required to reproduce an asymptotically linear nucleon spectrum in the ED context. Note that $m_5$ in this equation denotes the mass of the Dirac field in the asymptotic AdS$_5$ space. It is related to the conformal dimension in four dimensions via \cite{Iqbal:2009fd, Henningson:1998cd}
\begin{equation}
\Delta = 2 + m_5,\, .
\end{equation}
Here, we consider two cases: the canonical conformal dimensions for nucleons $\Delta=9/2$ with $m_5=5/2$, as well as an alternative value $\Delta=7/2$ with $m_5=3/2$, assuming an anomalous dimension contribution. The latter choice is motivated by the successful phenomenological description of the nucleon spectrum reported in Ref.~\cite{Abidin:2009hr}.

From these equations, we are going to obtain the spectrum and wave functions for nucleons in ED and SD gravities in the following sections.


\section{Nucleon spectra and wave functions in ED model}\label{modelI}

\subsection{Dependence of the nucleon spectra on $\lambda$ in the ED model}\label{einsteindilatonlambdavarying}

\begin{figure}[htp!]%
    \centering
    {{\includegraphics[width=7.2cm]{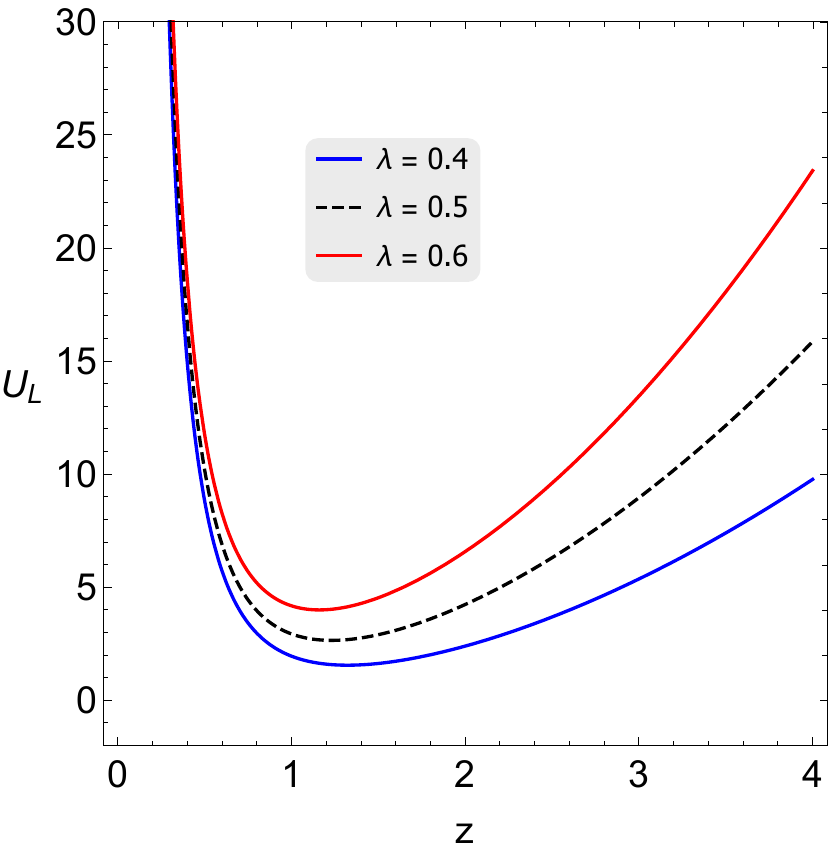}}
    \hskip 0.3cm 
    {\includegraphics[width=7.2cm]{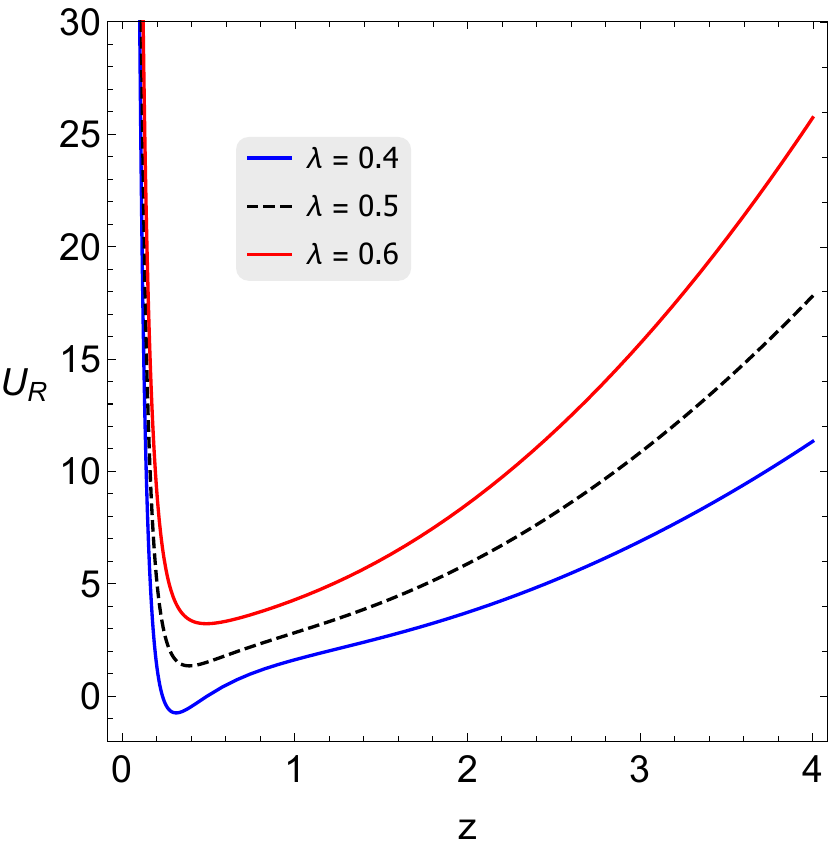}}}%
  \\ 
    \centering
    {{\includegraphics[width=7.2cm]{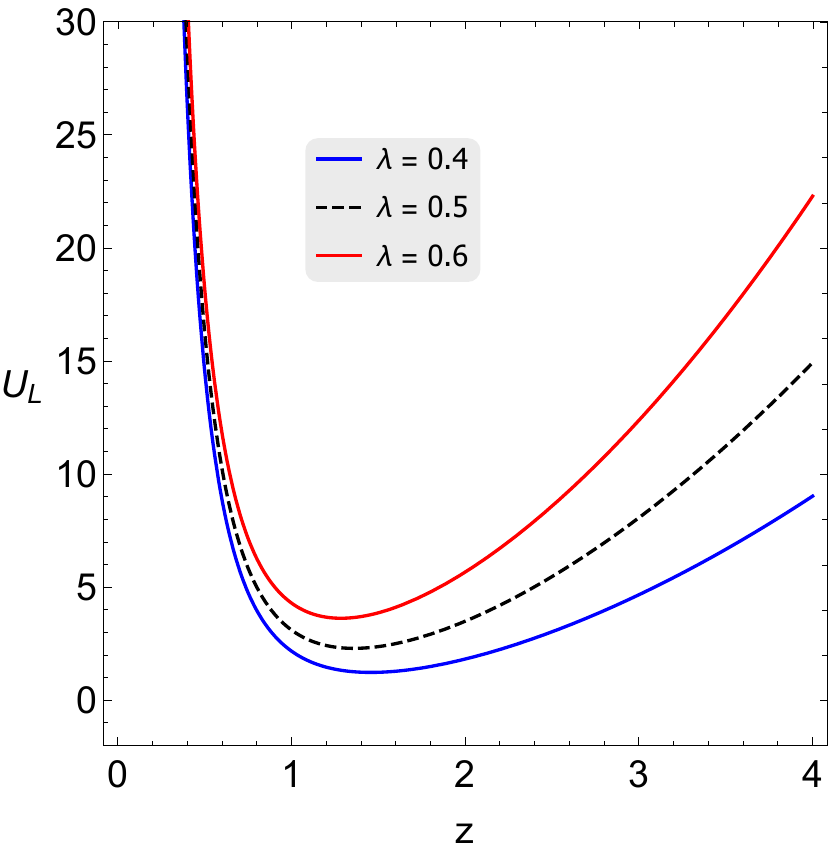}}
    \hskip 0.3cm 
    {\includegraphics[width=7.2cm]{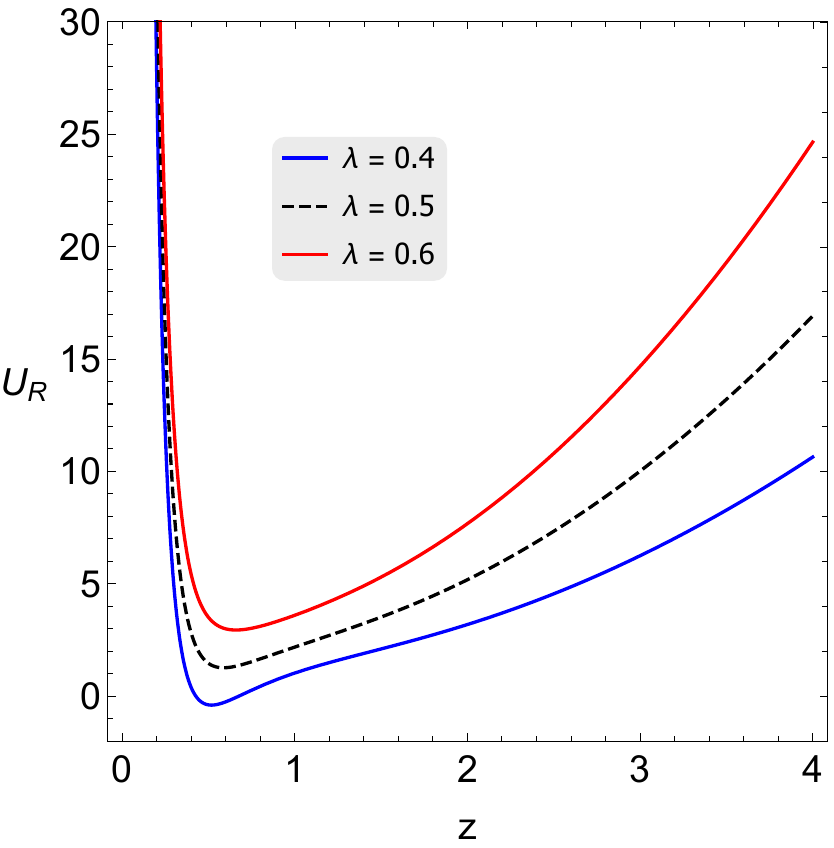}}}%
    
    \caption{Schrödinger potentials in ED holographic models as a function of coordinate $z$  and $\lambda = 0.4, 0.5, 0.6$. {\bf Upper panels}: $\Delta = 7/2$. {\bf Lower panels}: $\Delta = 9/2$. {\bf Left panels}: $U_{L}(z,\lambda)$. {\bf Right panels}: $U_{R}(z,\lambda)$.}%
    \label{Plot:NucleonsURLEDm52}%
\end{figure}

In this section, we describe the nucleon spectrum in ED gravity with the $\lambda$ variable. The choice $\lambda = 0.5$ reproduces the previous results reported in Ref.~\cite{Ballon-Bayona:2024yuz}. In addition, we will analyze the cases $\lambda = 0.4$ and $\lambda = 0.6$. We will adopt the conformal dimensions $\Delta = 7/2$ and $\Delta = 9/2$ for the three cases. The left- and right-handed components of the Schrödinger potential for nucleons are shown in Fig.~\ref{Plot:NucleonsURLEDm52} ($\Delta = 7/2$ and $\Delta = 9/2$) for each value of $\lambda$. The left panel displays a smoother potential, while the right panel exhibits a slight irregularity near the minimum for both values of the conformal dimension. In both cases, the minima are ordered such that $\lambda = 0.4$ corresponds to the lowest minimum, $\lambda = 0.5$ to an intermediate minimum, and $\lambda = 0.6$ to the highest minimum.

Near the boundary, the Schrödinger potentials for nucleons admit the expansion:
\begin{align}
   \nonumber  U_{R/L}(z, \lambda) =&\ \dfrac{(\Delta - 2)\left((\Delta - 2) \mp 1\right)}{z^{2}}  + \dfrac{6\,\lambda\left(\Delta - 2\right) - 4(\Delta - 2)^{2}}{z} + \dfrac{20(\Delta - 2)^{2}}{3} \\
    &+9\,\lambda^{2} \pm \dfrac{4(\Delta - 2)}{3} - 12\lambda(\Delta - 2)\,,
\end{align}
where we consider $k = 1$, for simplicity. On the other hand, for large $z$, the potentials behave as
\begin{align}
     U_{R/L}(z\to \infty, \lambda) = 4\,\lambda^{2}\,z^{2}\,.\label{approach2}
\end{align} 
Note that the result \eqref{approach2} does not depend on the parameter $\Delta$.  This behavior follows directly from the fact that $\lambda$ enters only the first term of the effective mass appearing in the Schrödinger potential, as shown in Eq.~\eqref{potentialed}.

In order to compute the nucleon spectra, we consider the values of the parameters $\lambda$, $\Delta$, and $k$ listed in Table \ref{Table:kmlambdanucleons}. For each choice of $\Delta$ and $\lambda$, the corresponding value of $k$ is fixed to reproduce the nucleon mass in the ground state correctly. As $\lambda$ decreases, the corresponding value of $k$ increases. 

\begin{table}[htp!]
\centering
\footnotesize
\begin{tabular}{c | c|c | c|c | c|c}
\hline\hline
$\lambda$
  & \multicolumn{2}{c|}{$\lambda = 0.6$}
  & \multicolumn{2}{c|}{$\lambda = 0.5$}
  & \multicolumn{2}{c}{$\lambda = 0.4$} \\
\cline{2-7}
\hline
$\Delta$
  & ${7}/{2}$ & ${9}/{2}$
  & ${7}/{2}$ & ${9}/{2}$
  & ${7}/{2}$ & ${9}/{2}$ \\
  \hline
  $k$ 
  & $(0.364\,\text{GeV})^{2}$ & $(0.375\,\text{GeV})^{2}$
  & $(0.425\,\text{GeV})^{2}$ & $(0.441\,\text{GeV})^{2}$
  & $(0.512\,\text{GeV})^{2}$ & $(0.539\,\text{GeV})^{2}$\\
\hline\hline
\end{tabular}
\caption{The values of the parameters $\lambda$, $\Delta$, and $k$ are fixed to reproduce the nucleon mass in the ground state.}
\label{Table:kmlambdanucleons}
\end{table}

 We also computed the relative error of the nucleon mass using the following expression:
\begin{equation}
\% m = \frac{|m_{\text{th}} - m_{\text{exp}}|}{m_{\text{exp}}}\times 100 \, ,
\end{equation}
where $\% m$ denotes the relative error of the mass, $m_{\text{th}}$ is the theoretical prediction, and $m_{\text{exp}}$ is the corresponding experimental value.

Our results for the nucleon spectrum and relative errors are presented in Table \ref{Table:NucleonsEDmodelm52notes} ($\Delta = 7/2$ and $\Delta = 9/2$) for $\lambda = 0.6, 0.5, 0.4$ in comparison with experimental data and the soft-wall model\footnote{The soft-wall model yields a good description of the nucleon spectrum for $\Delta = 5/2$, in good agreement with the experimental data. Similar values of the conformal dimension have also been employed in Ref.~\cite{Braga:2011wa}.}. Note that the spectrum of nucleons is sensitive to both conformal dimension and variation in $\lambda$. The nucleon spectrum increases as $\lambda$ decreases. Our results show that $\lambda = 0.6$ yields nucleon masses that deviate more significantly from the experimental values, particularly for the excited states. In contrast, the choice $\lambda = 0.4$ provides a better overall agreement with the experimental data, especially for higher radial excitations, indicating that smaller values of $\lambda$ improve the phenomenological description of the nucleon spectrum. The result with the smallest relative error, consequently, better agreement with experimental data is $1.3\%$ associated with the excited state $m_{N^{3}}$ for $\lambda = 0.4$.

\begin{table}[htp!]
\centering
\footnotesize
\begin{tabular}{c|c|cc|cc|cc|cc|c}
\hline\hline

$\Delta$
&
States
&
\multicolumn{2}{c|}{$\lambda=0.6$}
&
\multicolumn{2}{c|}{$\lambda=0.5$ \cite{Ballon-Bayona:2024yuz}}
&
\multicolumn{2}{c|}{$\lambda=0.4$}
&
\multicolumn{2}{c|}{Soft-wall \cite{Abidin:2009hr}}
&
Experimental
\\

\cline{3-10}

&
&
$m_{N^n}$ & $\%m$
&
$m_{N^n}$ & $\%m$
&
$m_{N^n}$ & $\%m$
&
$m_{N^n}$ & $\%m$
&
\\

\hline

&
$m_{N^0}$
&
0.938 & 0.0
&
0.938 & 0.0
&
0.938 & 0.0
&
0.938 & 0.0
&
$0.938 \pm 0.001$
\\

7/2
&
$m_{N^1}$
&
1.235 & 14.2
&
1.270 & 11.8
&
1.317 & 8.5
&
1.149 & 20.21
&
$1.440 \pm 0.030$
\\

&
$m_{N^2}$
&
1.472 & 13.9
&
1.531 & 10.5
&
1.608 & 6.0
&
1.327 & 22.40
&
$1.710 \pm 0.030$
\\

&
$m_{N^3}$
&
1.676 & 10.9
&
1.753 & 6.8
&
1.853 & 1.3
&
1.483 & 21.12
&
$1.880 \pm 0.050$
\\

\hline\hline

&
$m_{N^0}$
&
0.938 & 0.0
&
0.938 & 0.0
&
0.938 & 0.0
&
0.938 & 0.0
&
$0.938 \pm 0.001$
\\

9/2
&
$m_{N^1}$
&
1.252 & 13.06
&
1.295 & 10.07
&
1.354 & 5.97
&
1.083 & 24.79
&
$1.440 \pm 0.030$
\\

&
$m_{N^2}$
&
1.501 & 12.22
&
1.572 & 8.07
&
1.668 & 2.46
&
1.211 & 29.18
&
$1.710 \pm 0.030$
\\

&
$m_{N^3}$
&
1.714 & 8.83
&
1.806 & 3.93
&
1.931 & 2.71
&
1.327 & 29.41
&
$1.880 \pm 0.050$
\\

\hline\hline
\end{tabular}

\caption{
Nucleon spectrum in GeV for the cases $\Delta=7/2,9/2$ in ED gravity with $\lambda=0.4,0.5,0.6$ compared with experimental data \cite{ParticleDataGroup:2026aaa}.
}
\label{Table:NucleonsEDmodelm52notes}
\end{table}

\begin{figure}[htp!]%
    \centering
    {{\includegraphics[width=7.2cm]{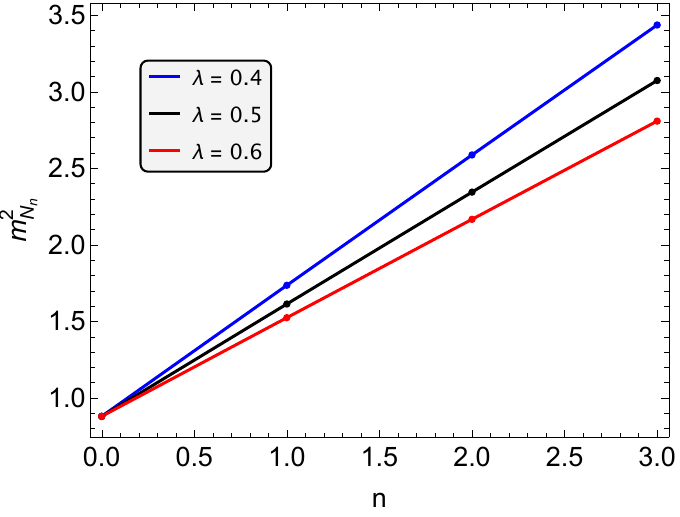}}
    \hskip 0.3cm 
    {\includegraphics[width=7.2cm]{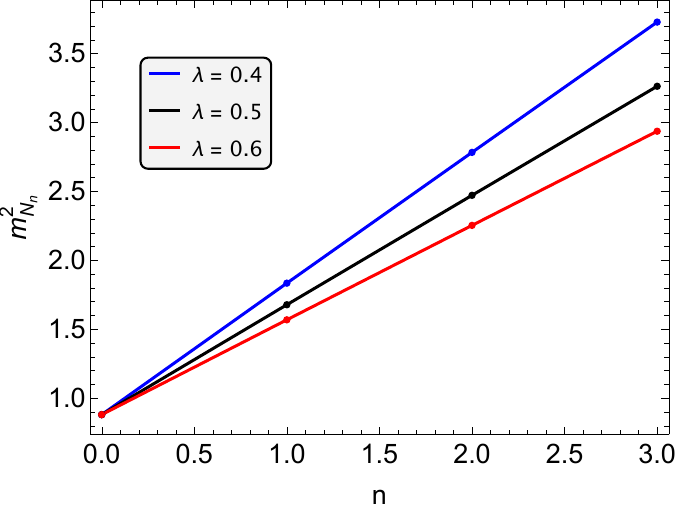}}}%
    
    \caption{Regge trajectories for nucleons considering the ground state and three radial excitations $n = 1, 2,3$ for $\lambda = 0.4, 0.5, 0.6$. {\bf Left panel}: $\Delta = 7/2$. 
 {\bf Right panel}: $\Delta = 9/2$.}%
    \label{Plot:ReggeNucleons}%
\end{figure}

In Table \ref{Table:ReggetrajectoriesEDlambdavarying} we present the Regge trajectories for each $\lambda$. The slope increases progressively from $\lambda = 0.6$ to $\lambda = 0.4$ for both conformal dimension values. The slope confirm that the nucleon spectrum for $\lambda = 0.6$ is smallest in comparison with $\lambda = 0.5$ and  $\lambda = 0.4$ .
\begin{table}[htp!]
\centering
\begin{tabular}{c|c|c|c}
\hline 
\hline
$\Delta$  & $\lambda = 0.6$ &$\lambda = 0.5$  & $\lambda = 0.4$ \\
\hline 
7/2  & $0.642\,n + 0.882$ & $0.729\,n + 0.883$ & $0.849\,n + 0.884$\\
\hline\hline
9/2 &  $0.684\,n + 0.883$&  $0.904\,n + 0.884$ & $0.946\,n + 0.886$\\
\hline\hline
\end{tabular}
\caption{Regge trajectories for nucleons for $\Delta = 7/2, 9/2$ and  $\lambda = 0.4, 0.5, 0.6$.}
\label{Table:ReggetrajectoriesEDlambdavarying}
\end{table}

In Fig.~\ref{Plot:ReggeNucleons}, we present the Regge trajectories of nucleons for $\lambda = 0.4$ (blue line), $\lambda = 0.5$ (black line) and $\lambda = 0.6$ (red line), considering $\Delta = 7/2$ (left panel) and  $\Delta = 9/2$ (right panel). In accordance with the results shown in the previous tables above, the Regge trajectories of the nucleons exhibit a clear dependence on the parameter $\lambda$. For both cases, ($\Delta=7/2$) and  ($\Delta=9/2$), the trajectory corresponding to $\lambda = 0.4$ displays the largest slope, while increasing $\lambda$ leads to progressively smaller slopes, with $\lambda = 0.6$ yielding the smallest slope. This behavior indicates that smaller values of $\lambda$ enhance the radial excitation, resulting in a faster growth of $m_n^2$ with the radial quantum number $n$.

Based on the nucleon spectra presented in Table \ref{Table:NucleonsEDmodelm52notes}, in the next subsection, we present the corresponding wave functions for each value of $\lambda$ and  conformal dimension.


\subsection{Wave functions in ED model}

\begin{figure}[htp!]%
    \centering
    {{\includegraphics[width=7.2cm]{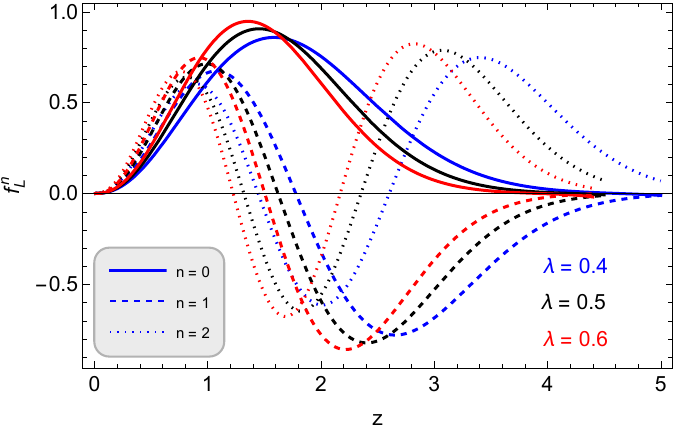}}
    \hskip 0.3cm 
    {\includegraphics[width=7.2cm]{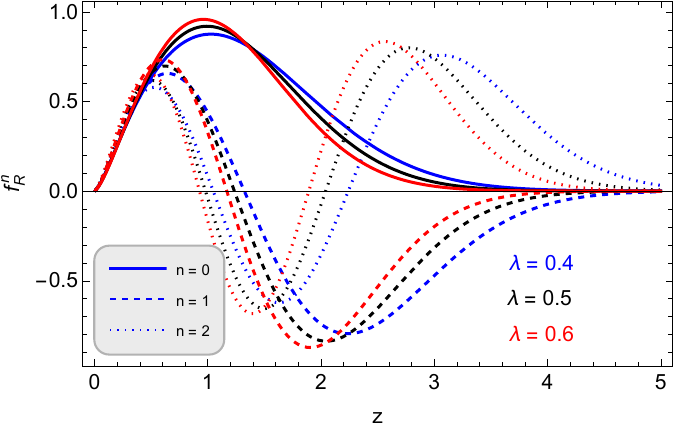}}}%
\\
\vskip0.5cm 
    \centering
    {{\includegraphics[width=7.2cm]{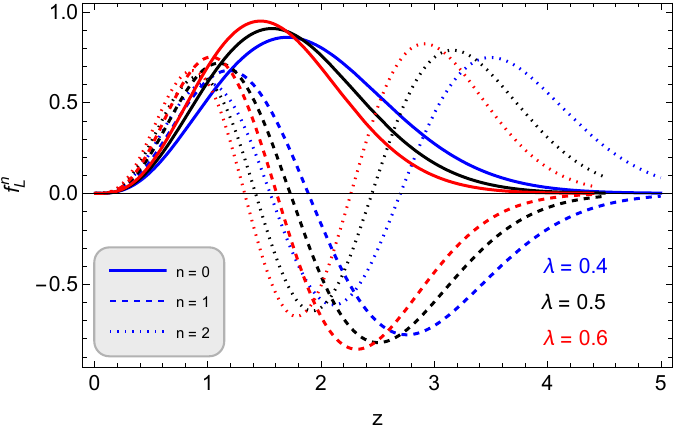}}}
    \hskip 0.3cm 
    {\includegraphics[width=7.2cm]{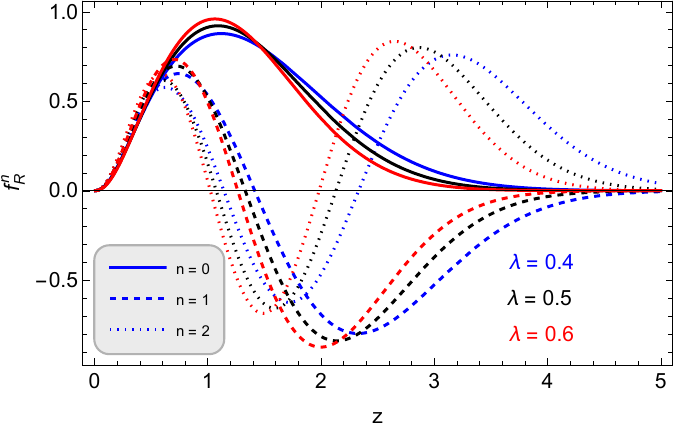}}%
    
    \caption{Nucleon wave functions in ED holographic model for $n = 0,1,2$ and $\lambda = 0.4, 0.5, 0.6$.    
    {\bf Upper left panel}: $f_{L}^{n}$ with $\Delta = 7/2$. 
{\bf Upper right panel}: $f_{R}^{n}$ with $\Delta = 7/2$. {\bf Lower left panel}: $f_{L}^{n}$ with $\Delta = 9/2$.  
{\bf Lower right panel}: $f_{R}^{n}$ with $\Delta = 9/2$.}%
    \label{Plot:WaveNucleonsm52}%
\end{figure}

Figure \ref{Plot:WaveNucleonsm52} shows the left and right nucleon wave functions as a function of the holographic coordinate $z$ for $\lambda = 0.4, 0.5, 0.6$, considering $\Delta = 7/2$ and $\Delta = 9/2$ for each $\lambda$ value. Observe that increasing $\lambda$ systematically shifts the positions of the extrema and alters the depth of the minima. The higher values of $\lambda$ tend to modify their oscillatory behavior along the holographic coordinate $z$. These features indicate a clear sensitivity of the nucleon wave functions to the parameter $\lambda$.  Note that the red curves ($\lambda = 0.6$) present the largest extrema, both in terms of peak heights and the depth of the minima, when compared with the black ($\lambda = 0.5$) and blue ($\lambda = 0.4$) curves. The black curves display an intermediate behavior, lying between the blue and red ones, while the red curves exhibit the smallest extrema. This ordered pattern reflects a clear hierarchy among modes and is directly related to the corresponding mass spectrum, with higher values of $\lambda$ leading to more pronounced wave-function profiles.

In the following subsection, we will adopt a complementary approach by studying the nucleon spectrum within the framework of SD gravity. This analysis aims to determine whether the discrepancies with respect to the experimental data for $\lambda = 0.6$ and $\lambda = 0.4$ can be reduced.

\section{Nucleon spectra in SD model}\label{modelII}

\subsection{Nucleon spectra in SD model A}\label{spectrumsd}
In this subsection we investigate the contributions due to SD gravity to the nucleon spectrum with $\lambda = 0.5$ fixed. For this reason, we fix $\lambda=0.5$ and vary the parameter $\alpha$, considering the cases $\Delta = 7/2$ and $\Delta = 9/2$, in order to improve the description of the nucleon spectrum. As presented in subsection~\ref{scalefactordilatonfield}, corrections controlled by the parameter $\alpha$ modify both the dilaton field and the scale factor when expressed in the string frame. These modifications, in turn, affect the effective mass $M_{\text{eff}}$, the Schrödinger potentials $U_{R/L}$ in Eq.~\eqref{potentialed}, and the corresponding eigenfunctions $\Psi_{R/L}$ in Eq.~\eqref{nucleonschrodingereqed}. Consequently, these quantities in the context of SD gravity will be written as
\begin{align}
    M_{\text{eff}} &= M_{\text{eff}}(z,k, \alpha), \\
    U_{R/L} &= U_{R/L}(z,k ,\alpha).\label{potentialsd}
\end{align}

\begin{figure}[htp!]%
    \centering
    {{\includegraphics[width=7.2cm]{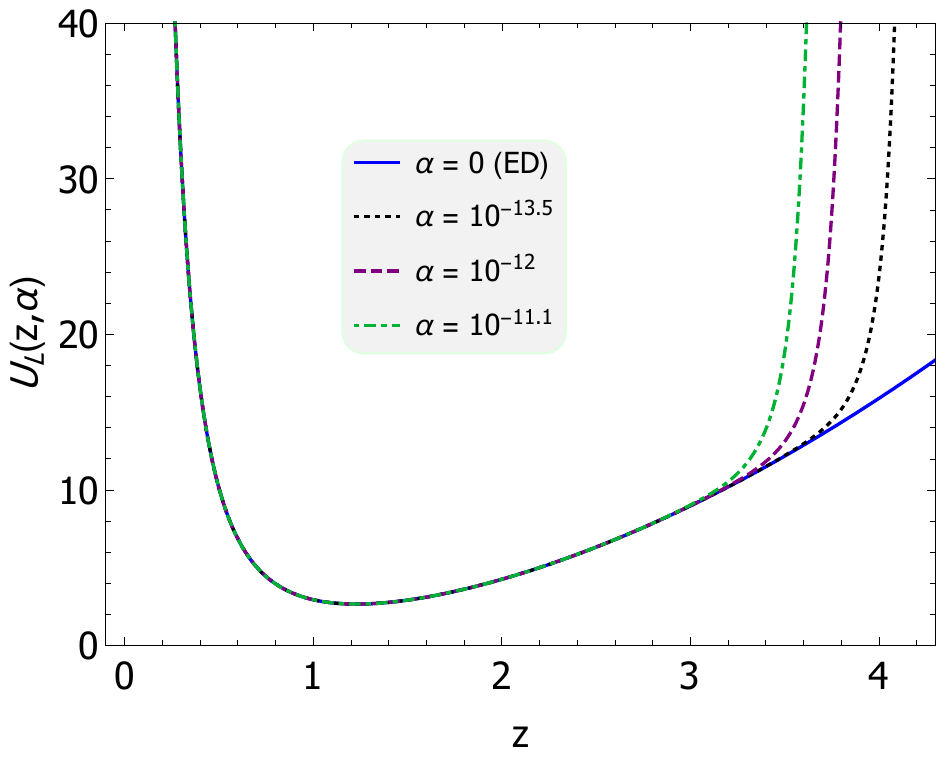}}
    \hskip 0.3cm
    {\includegraphics[width=7.2cm]{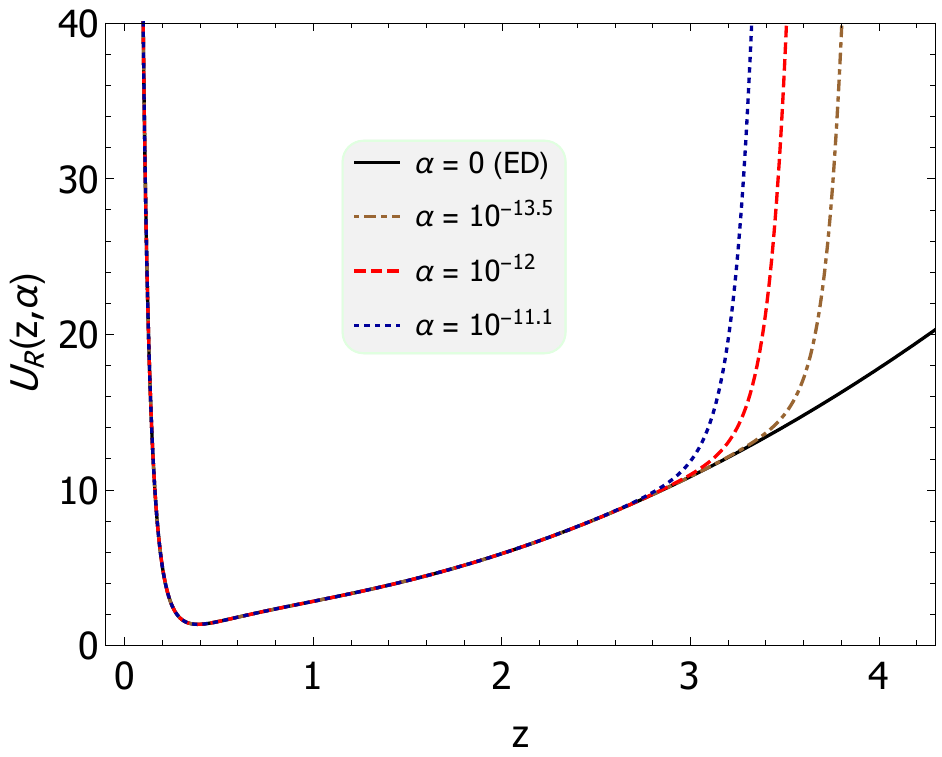}}}%
  \\
  \vskip 0.3cm 
    \centering
    {{\includegraphics[width=7.2cm]{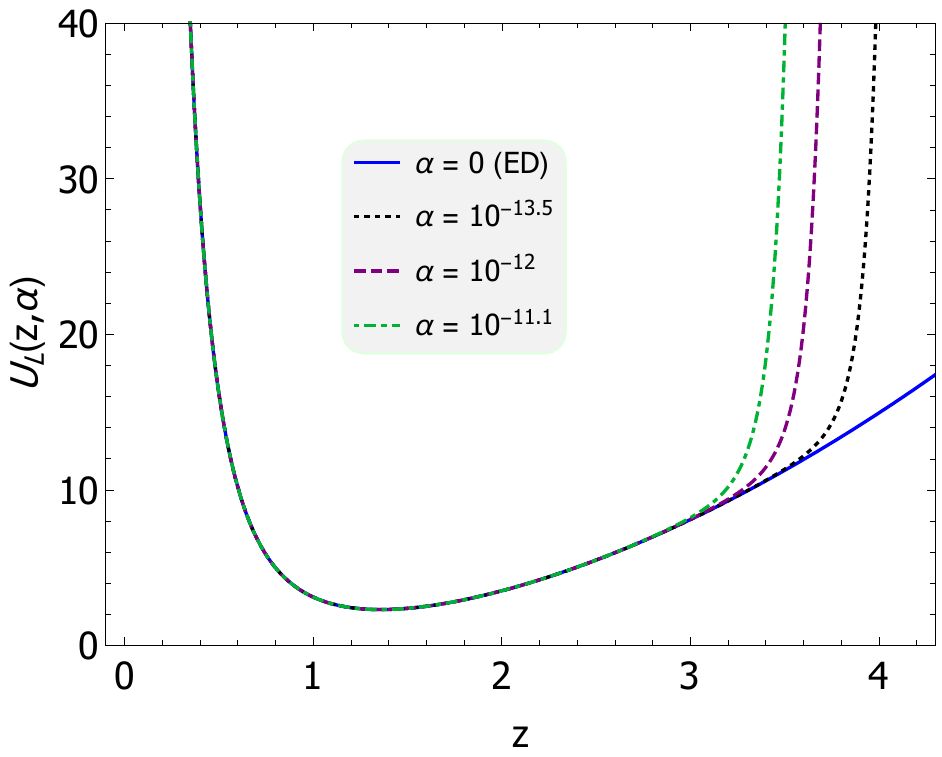}}
    \hskip 0.3cm
    {\includegraphics[width=7.2cm]{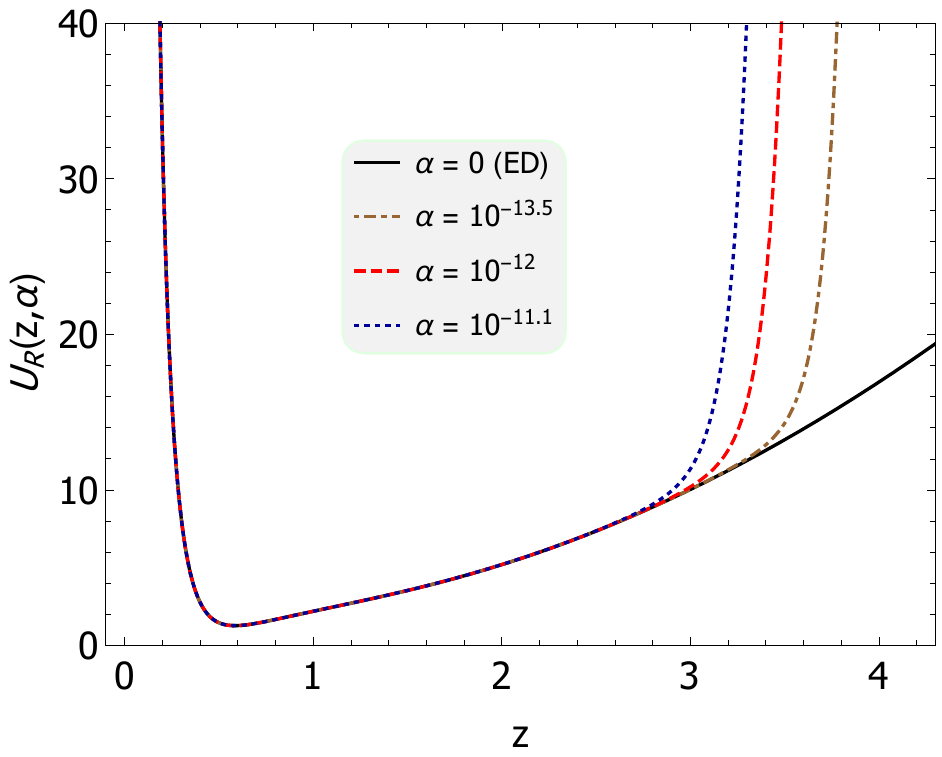}}}%
    
    \caption{Schrödinger potentials in ED $(\alpha=0)$ and SD ($\alpha = 10^{-13.5}, 10^{-12}, 10^{-11.1}$) gravity models as a function of coordinate $z$   for  $\lambda= 0.5$.  {\bf Upper panels}: $\Delta = 7/2$.  {\bf Lower panels}: $\Delta = 9/2$.
    {\bf Left panels}: $U_{L}(z,\alpha)$. 
 {\bf Right panels}: $U_{R}(z,\alpha)$.}%
    \label{Plot:URLEDm52Starobinsky}%
\end{figure}

The left- and right-handed components of the Schrödinger potentials for $\alpha = 0$, corresponding to pure ED gravity, and for $\alpha = 10^{-13.5}$, $10^{-12}$, and $10^{-11.1}$, which represent the SD scenario, are displayed in Fig.~\ref{Plot:URLEDm52Starobinsky} for $\Delta = 7/2$ and  $\Delta = 9/2$. These values of $\alpha$ are chosen to analyze the impact of Starobinsky corrections on the nucleon spectrum. 
The first nontrivial deviations from the ED results appear around $\alpha = 10^{-13.5}$, whereas $\alpha = 10^{-11.1}$ corresponds to the case where the effects are most pronounced while the dilaton profile remains stable. The fractional powers in the exponent are chosen to keep the system within the stability region while still allowing significant Starobinsky corrections. Observe that the ultraviolet behavior remains essentially unchanged, indicating that Starobinsky corrections do not contribute in this regime and the ED dominates. In contrast, Starobinsky corrections induce an abrupt rise in the Schrödinger potential, particularly in the infrared region. As the parameter $\alpha$ increases, these corrections become more pronounced, leading to a progressive enhancement of the Schrödinger potential and indicating a stronger confinement effect.

Near the boundary, the Schrödinger potential takes the form
\begin{align}
    U_{R/L}(z, \alpha) &= \dfrac{(\Delta - 2)\left((\Delta - 2) \mp 1\right)}{z^{2}} 
    + \dfrac{(\Delta - 2)\left(3 - 4\Delta + 8)\right)\sqrt{1 + 8\alpha}}{z} 
    + \dfrac{9}{4} \cr 
    & \;\quad - \dfrac{(18 \pm 4 + 20\Delta - 40)(\Delta - 2)}{3}  
    + (42 \pm 24)\alpha  
    + \cdots ,
\label{nucleonspotentialstarobinsky}
\end{align}
where we fixed $k = 1$ to better highlight the $\alpha$-dependent corrections to the Schrödinger potential. The first term in Eq.~\eqref{nucleonspotentialstarobinsky} corresponds to the pure $\rm AdS_{5}$ contribution. The second term contains the first $\alpha$ correction to this Schrödinger potential.

\begin{table}[htp!]
\centering
\resizebox{\textwidth}{!}{
\begin{tabular}{c|c|cc|cc|cc|cc|cc|c}
\hline\hline

$\Delta$
&
States
&
\multicolumn{2}{c|}{$\alpha=0$}
&
\multicolumn{2}{c|}{$10^{-13.5}$}
&
\multicolumn{2}{c|}{$10^{-12}$}
&
\multicolumn{2}{c|}{$10^{-11.1}$}
&
\multicolumn{2}{c|}{Soft-wall \cite{Abidin:2009hr}}
&
Experimental
\\

\cline{3-12}

&
&
$m_{N}$ & $\%m$
&
$m_{N}$ & $\%m$
&
$m_{N}$ & $\%m$
&
$m_{N}$ & $\%m$
&
$m_{N}$ & $\%m$
&
\\

\hline

&
$m_{N^0}$
&
0.938 & 0.0
&
0.938 & 0.0
&
0.938 & 0.0
&
0.938 & 0.0
&
0.938 & 0.0
&
$0.938 \pm 0.001$
\\

7/2
&
$m_{N^1}$
&
1.270 & 11.81
&
1.270 & 11.81
&
1.273 & 11.60
&
1.277 & 11.32
&
1.149 & 20.21
&
$1.440 \pm 0.030$
\\

&
$m_{N^2}$
&
1.531 & 10.47
&
1.534 & 10.29
&
1.557 & 8.95
&
1.577 & 7.78
&
1.327 & 22.40
&
$1.710 \pm 0.030$
\\

&
$m_{N^3}$
&
1.753 & 6.76
&
1.778 & 5.43
&
1.849 & 1.65
&
1.895 & 0.8
&
1.483 & 21.12
&
$1.880 \pm 0.050$
\\

\hline\hline

&
$m_{N^0}$
&
0.938 & 0.0
&
0.938 & 0.0
&
0.938 & 0.0
&
0.939 & 0.0
&
0.938 & 0.0
&
$0.938 \pm 0.001$
\\

9/2
&
$m_{N^1}$
&
1.295 & 10.07
&
1.295 & 10.07
&
1.300 & 9.72
&
1.306 & 9.31
&
1.083 & 24.79
&
$1.440 \pm 0.030$
\\

&
$m_{N^2}$
&
1.572 & 8.07
&
1.573 & 8.01
&
1.612 & 5.73
&
1.640 & 4.09
&
1.211 & 29.18
&
$1.710 \pm 0.030$
\\

&
$m_{N^3}$
&
1.806 & 3.93
&
1.827 & 2.82
&
1.936 & 2.98
&
1.992 & 5.96
&
1.327 & 29.41
&
$1.880 \pm 0.050$
\\

\hline\hline
\end{tabular}
}
\caption{
Nucleon masses in GeV in ED ($\alpha=0$) and
SD ($\alpha = 10^{-13.5},\,10^{-12},\,10^{-11.1}$)
holographic models with $\lambda = 0.5$, considering the ground state and
three radial excitations, $n=1,2,3$, for $\Delta = 7/2,9/2$ compared against
experimental data \cite{ParticleDataGroup:2026aaa}.
}
\label{Table:NucleonsMassesm52starobinsky}
\end{table}

Our results for the nucleon spectrum in SD gravity are presented in Table \ref{Table:NucleonsMassesm52starobinsky} corresponding to $\Delta = 7/2$ and $\Delta = 9/2$. We consider the ED case ($\alpha = 0$) and the SD scenarios with $\alpha = 10^{-13.5}$, $\alpha = 10^{-12}$, and $\alpha = 10^{-11.1}$, and compare the resulting spectra with the available experimental data and soft-wall model. Since corrections in $\alpha$ strengthen confinement of the Schrödinger potential far from the boundary, according to Fig.~\ref{Plot:URLEDm52Starobinsky}, this directly impacts the nucleon spectrum in the most excited states. As shown in Table~\ref{Table:NucleonsMassesm52starobinsky}, the ground-state mass $m_{N^{0}}$ remains essentially unchanged as the parameter $\alpha$ increases. In contrast, the effects of Starobinsky corrections become apparent already at the level of the first radial excitation, $m_{N^{1}}$, and grow progressively stronger for the second and third excited states, $m_{N^{2}}$ and $m_{N^{3}}$, respectively, thereby bringing the nucleon spectrum closer to the experimental data. On the other hand, the results for the spectrum of nucleons described in the soft-wall model are discrepant compared to the data. We compute the relative errors of the nucleon spectrum with respect to the experimental data, as presented in Table~\ref{Table:NucleonsMassesm52starobinsky}. The smallest relative error is $0.8\%$, corresponding to the third excited state, $m_{N^{3}}$, for $\Delta = 7/2$.

\begin{figure}[htp!]%
    \centering
    {{\includegraphics[width=7.2cm]{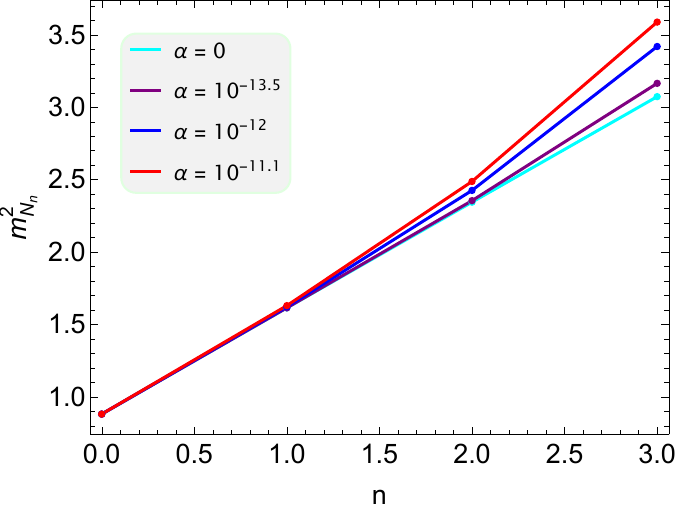}}
    \hskip 0.3cm 
    {\includegraphics[width=7.2cm]{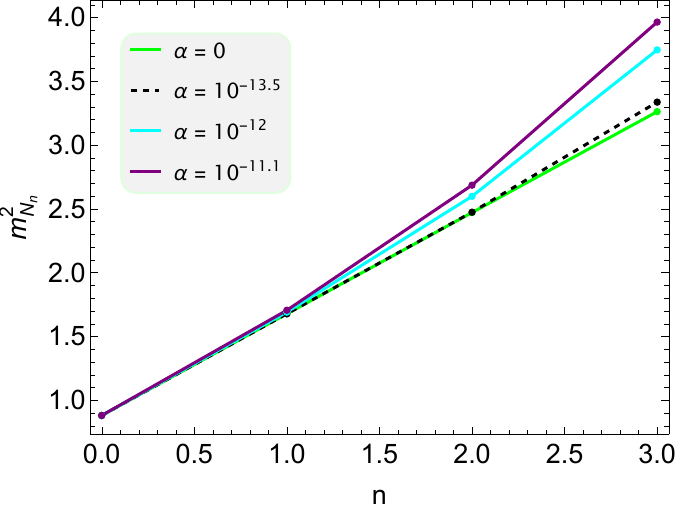}}}%
    
    \caption{Squared nucleon masses as a function of the radial excitation number $n$ in SD holographic models, considering the ground state and the first three radial excitations $n = 1, 2,3$ for $\alpha = 0$, $\alpha = 10^{-13.5}$, $\alpha = 10^{-12}$ and $\alpha = 10^{-11.1}$. {\bf Left panel}: $\Delta = 7/2$.
 {\bf Right panel}: $\Delta = 9/2$.}%
    \label{Plot:spectrumnucleonsStarobinsky}%
\end{figure}

In Fig.~\ref{Plot:spectrumnucleonsStarobinsky}, we display the squared nucleon masses $m_{N^{n}}^{2}$ as a function of the radial excitation number $n$, for $\alpha = 0$ (ED context), $\alpha = 10^{-13.5}$, $\alpha = 10^{-12}$, and $\alpha = 10^{-11.1}$, considering $m = 3/2$ (left panel) and $m = 5/2$ (right panel). In the ED case, the spectrum exhibits an asymptotically linear behavior, while in the SD gravity scenario, it becomes non-linear due to the $\alpha$-dependent corrections. Non-linearity in the spectrum has already been observed in the hyperons $\Sigma$ \cite{Guo:2024nrf} and mesons \cite{MartinContreras:2020cyg}.
 
To quantify the deviations from linear Regge trajectories induced by the $\alpha$ corrections, we fit the nucleon spectrum using the phenomenological parametrization following Ref.~\cite{MartinContreras:2020cyg}, namely
\begin{align}
m_{N^{n}}^{2} = a\,(n + b)^{\nu},
\end{align}
where $\nu=1$ corresponds to a linear Regge trajectory, whereas $\nu>1$ characterizes a nonlinear behavior. In Table \ref{Table:ReggetrajectoriesStarobinskylambda05}, we present the parameters obtained from the phenomenological fit to the nucleon Regge trajectories, highlighting the deviations from linearity induced by the $\alpha$ corrections. As $\alpha$ increases, the fitted exponent $\nu$ deviates further from unity, signaling a progressively stronger departure from linear Regge behavior. This trend is fully consistent with the behavior observed in Fig.~\ref{Plot:spectrumnucleonsStarobinsky}.

\begin{table}[htp!]
\centering
\begin{tabular}{c|c|c|c|c}
\hline 
\hline
$\Delta$  & $\alpha = 0$ & $\alpha = 10^{-13.5}$ & $\alpha = 10^{-12}$ & $\alpha = 10^{- 11.1}$\\
\hline 
7/2  & $0.729\,n + 0.883$ & $0.576(n + 1.45)^{1.14}$ &  $0.268(n + 2.16)^{1.55}$ &  $0.161(n + 2.58)^{1.80}$\\
\hline\hline
9/2 &  $0.904\,n + 0.884$ & $0.669(n + 1.29)^{1.10}$ & $0.249(n + 2.16)^{1.65}$  & $0.155(n + 2.51)^{1.90}$\\
\hline\hline
\end{tabular}
\caption{Phenomenological parametrization of nucleon Regge trajectories in the ED and SDA models.}
\label{Table:ReggetrajectoriesStarobinskylambda05}
\end{table}


\subsection{Wave functions in SD model A}

\begin{figure}[htp!]%
    \centering
    {{\includegraphics[width=7.2cm]{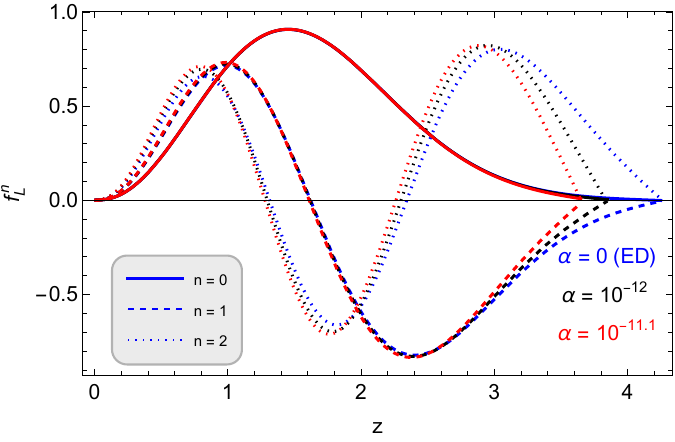}}
    \hskip 0.3cm
    {\includegraphics[width=7.2cm]{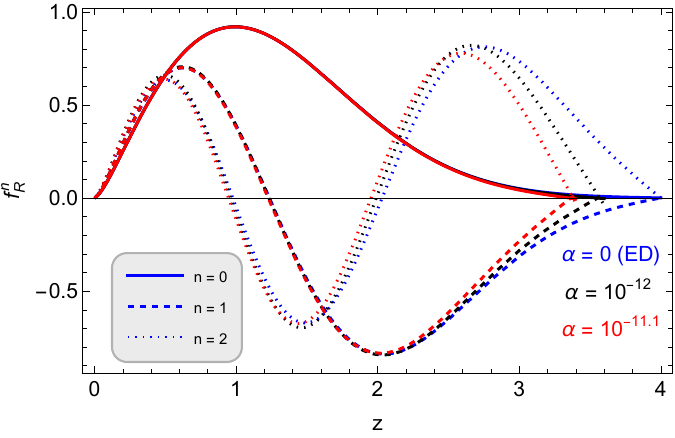}}}%
   \\ 
    \centering
    {{\includegraphics[width=7.2cm]{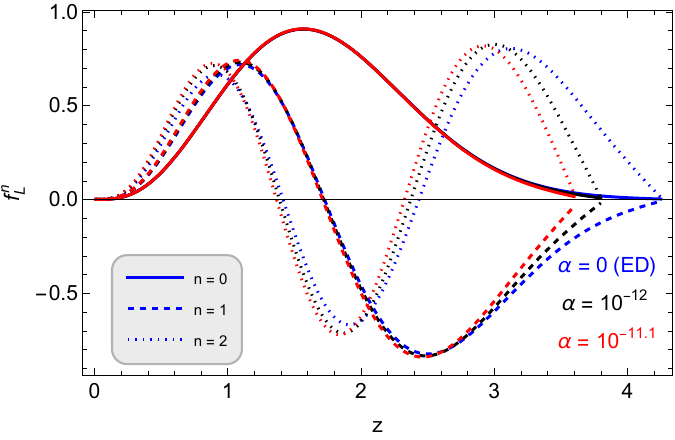}}
    \hskip 0.3cm 
    {\includegraphics[width=7.2cm]{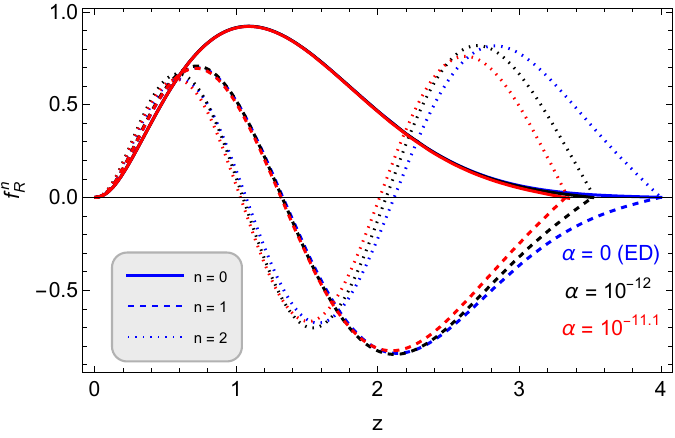}}}%
    
    \caption{Nucleon wave functions in ED $(\alpha=0)$ and SD $(\alpha=10^{-12}, 10^{-11.1})$ holographic models with $n = 0, 1, 2$ for $\lambda=0.5$. {\bf Upper panels}: $\Delta = 7/2$. {\bf Lower panels}: $\Delta = 9/2$. {\bf Left panels}: $f_{L}^{n}(z,\alpha)$. {\bf Right panels}: $f_{R}^{n}(z,\alpha)$.}%
    \label{Plot:WaveNucleonsm52starobinskykappa}%
\end{figure}

The behavior of the nucleon wave functions in ED gravity, corresponding to $\alpha = 0$, and in SD model A, for $\alpha = 10^{-12}$ and $\alpha = 10^{-11.1}$, for the ground state of the nucleon and first and second excited states, i.e., $n = 0,1,2$, are displayed in Fig.~\ref{Plot:WaveNucleonsm52starobinskykappa}. It is observed that small deviations emerge in the nucleon wave functions as the parameter $\alpha$ increases, compared to the standard $\alpha = 0$ case. As expected, the effects of SD gravity on the nucleon wave functions are more pronounced for the state $m_{N^{2}}$, which exhibits two nodes, for both conformal dimensions $\Delta = 7/2$ and $\Delta = 9/2$. This behavior is consistent with the nucleon spectra presented in Table \ref{Table:NucleonsMassesm52starobinsky}.

\subsection{Nucleon spectra in SD model B}
\label{spectrumsdalternative}

In the present subsection, we will describe the corrections to the nucleon spectra within the framework of the Starobinsky model B as a complementary approach to the ED gravity model with a varying parameter $\lambda$ developed in the previous subsection \ref{einsteindilatonlambdavarying}. The nucleon spectrum obtained for $\lambda = 0.6$ and $\lambda = 0.4$ exhibits deviations from the experimental data. 
As shown in \ref{einsteindilatonlambdavarying}, for $\lambda = 0.6$ the relative errors range from $14.2\%$ to $10.9\%$ for $\Delta = 7/2$ and from $13.06\%$ to $8.83\%$ for $\Delta = 9/2$.  For $\lambda = 0.4$, the relative errors are reduced, ranging from $8.5\%$ to $1.3\%$ for $\Delta = 7/2$ and from $5.97\%$ to $2.71\%$ for $\Delta = 9/2$. In this context, we incorporate Starobinsky gravity to improve the nucleon spectrum for these cases. 

The behavior of the Schrödinger potential for $\lambda = 0.4$ and $\lambda = 0.6$ with $\alpha$ varying is similar to \eqref{nucleonspotentialstarobinsky},  differing only in some coefficients. 

The nucleon spectra for the SD model B, {\sl i.e.}, with varying $\lambda$ and $\alpha$, are presented in Tables \ref{Table:NucleonsMassesm52starobinskylambda06} and \ref{Table:NucleonsMassesm52starobinskylambda04} for $\Delta = 7/2$ and $\Delta = 9/2$, respectively. The results are shown for the ED case ($\alpha = 0$) and for the SD gravities with $\alpha = 10^{-13.5}$, $10^{-12}$, and $10^{-11.1}$, and are compared with the experimental data and the soft-wall model. We observe that the nucleon masses increase progressively with $\alpha$, bringing the theoretical predictions closer to the experimental data for some excited states, while exceeding them for others.

The corresponding relative errors are also listed in these tables. As expected, the $\alpha$ corrections predominantly enhance the masses of the excited states. For $\lambda = 0.6$, the smallest relative error is $3.78\%$, obtained for the third excited state, whereas for $\lambda = 0.4$ the best agreement is achieved for the second excited state, with a relative error of only $0.29\%$.

\vskip 0.5cm 
\begin{table}[htp!]
\centering
\resizebox{\textwidth}{!}{
\begin{tabular}{c|c|cc|cc|cc|cc|cc|c}
\hline\hline

$\Delta$
&
States
&
\multicolumn{2}{c|}{$\alpha=0$}
&
\multicolumn{2}{c|}{$10^{-13.5}$}
&
\multicolumn{2}{c|}{$10^{-12}$}
&
\multicolumn{2}{c|}{$10^{-11.1}$}
&
\multicolumn{2}{c|}{Soft-wall \cite{Abidin:2009hr}}
&
Experimental
\\

\cline{3-12}

&
&
$m_{N}$ & $\%m$
&
$m_{N}$ & $\%m$
&
$m_{N}$ & $\%m$
&
$m_{N}$ & $\%m$
&
$m_{N}$ & $\%m$
&
\\

\hline

&
$m_{N^0}$
&
0.938 & 0.0
&
0.938 & 0.0
&
0.938 & 0.0
&
0.938 & 0.0
&
0.938 & 0.0
&
$0.938 \pm 0.001$
\\

7/2
&
$m_{N^1}$
&
1.235 & 14.24
&
1.235 & 14.24
&
1.235 & 14.24
&
1.236 & 14.17
&
1.149 & 20.21
&
$1.440 \pm 0.030$
\\

&
$m_{N^2}$
&
1.472 & 13.2
&
1.473 & 13.86
&
1.479 & 13.52
&
1.489 & 12.92
&
1.327 & 22.40
&
$1.710 \pm 0.030$
\\

&
$m_{N^3}$
&
1.676 & 10.85
&
1.691 & 10.05
&
1.717 & 8.67
&
1.745 & 7.18
&
1.483 & 21.12
&
$1.880 \pm 0.050$
\\

\hline\hline

&
$m_{N^0}$
&
0.938 & 0.0
&
0.938 & 0.0
&
0.938 & 0.0
&
0.939 & 0.0
&
0.938 & 0.0
&
$0.938 \pm 0.001$
\\

9/2
&
$m_{N^1}$
&
1.252 & 13.06
&
1.252 & 13.06
&
1.253 & 12.99
&
1.255 & 12.85
&
1.083 & 24.79
&
$1.440 \pm 0.030$
\\

&
$m_{N^2}$
&
1.501 & 12.22
&
1.505 & 11.99
&
1.515 & 11.40
&
1.529 & 10.59
&
1.211 & 29.18
&
$1.710 \pm 0.030$
\\

&
$m_{N^3}$
&
1.714 & 8.83
&
1.734 & 7.77
&
1.773 & 5.69
&
1.809 & 3.78
&
1.327 & 29.41
&
$1.880 \pm 0.050$
\\

\hline\hline
\end{tabular}
}
\caption{Nucleon masses in GeV in ED $(\alpha=0)$ and SD ($\alpha = 10^{-13.5}, 10^{-12}, 10^{-11.1}$) holographic models considering the ground state and three radial excitations $n = 1, 2,3$ for $\Delta = 7/2, 9/2$ with $\lambda = 0.6$. For comparison, we also show the corresponding experimental data \cite{ParticleDataGroup:2026aaa}.}
\label{Table:NucleonsMassesm52starobinskylambda06}
\end{table}

\vskip 0.5cm 
\begin{table}[htp!]
\centering
\resizebox{\textwidth}{!}{
\begin{tabular}{c|c|cc|cc|cc|cc|cc|c}
\hline\hline

$\Delta$
&
States
&
\multicolumn{2}{c|}{$\alpha=0$}
&
\multicolumn{2}{c|}{$10^{-13.5}$}
&
\multicolumn{2}{c|}{$10^{-12}$}
&
\multicolumn{2}{c|}{$10^{-11.1}$}
&
\multicolumn{2}{c|}{Soft-wall \cite{Abidin:2009hr}}
&
Experimental
\\

\cline{3-12}

&
&
$m_{N}$ & $\%m$
&
$m_{N}$ & $\%m$
&
$m_{N}$ & $\%m$
&
$m_{N}$ & $\%m$
&
$m_{N}$ & $\%m$
&
\\

\hline

&
$m_{N^0}$
&
0.938 & 0.0
&
0.938 & 0.0
&
0.938 & 0.0
&
0.938 & 0.0
&
0.938 & 0.0
&
$0.938 \pm 0.001$
\\

7/2
&
$m_{N^1}$
&
1.317 & 8.54
&
1.318 & 8.40
&
1.338 & 7.08
&
1.349 & 6.32
&
1.149 & 20.21
&
$1.440 \pm 0.030$
\\

&
$m_{N^2}$
&
1.608 & 5.96
&
1.628 & 4.80
&
1.705 & 0.29
&
1.741 & 1.81
&
1.327 & 22.40
&
$1.710 \pm 0.030$
\\

&
$m_{N^3}$
&
1.853 & 1.44
&
1.941 & 3.24
&
2.087 & 11.01
&
2.148 & 14.26
&
1.483 & 21.12
&
$1.880 \pm 0.050$
\\

\hline\hline

&
$m_{N^0}$
&
0.938 & 0.0
&
0.938 & 0.0
&
0.938 & 0.0
&
0.939 & 0.0
&
0.938 & 0.0
&
$0.938 \pm 0.001$
\\

9/2
&
$m_{N^1}$
&
1.354 & 5.97
&
1.355 & 5.90
&
1.387 & 3.68
&
1.407 & 2.29
&
1.083 & 24.79
&
$1.440 \pm 0.030$
\\

&
$m_{N^2}$
&
1.668 & 2.46
&
1.684 & 1.52
&
1.807 & 5.67
&
1.858 & 8.65
&
1.211 & 29.18
&
$1.710 \pm 0.030$
\\

&
$m_{N^3}$
&
1.931 & 2.71
&
2.018 & 7.34
&
2.238 & 19.04
&
2.320 & 23.40
&
1.327 & 29.41
&
$1.880 \pm 0.050$
\\

\hline\hline
\end{tabular}
}
\caption{Nucleon masses in GeV in ED $(\alpha=0)$ and SD ($\alpha = 10^{-13.5}, 10^{-12}, 10^{-11.1}$) holographic models considering the ground state and three radial excitations $n = 1, 2,3$ for $\Delta = 7/2, 9/2$ with  $\lambda = 0.4$. For comparison, we also show the corresponding experimental data \cite{ParticleDataGroup:2026aaa}.}
\label{Table:NucleonsMassesm52starobinskylambda04}
\end{table}

\begin{figure}[htp!]%
    {{\includegraphics[width=7.45cm]{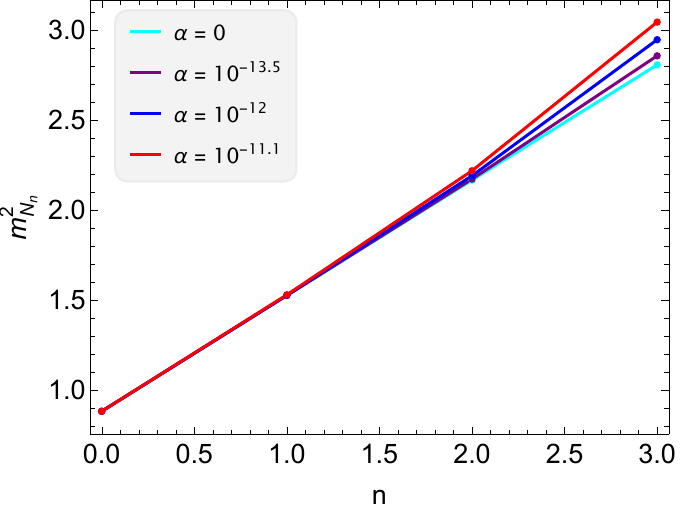}}
    \hskip 0.3cm 
    {\includegraphics[width=7.2cm]{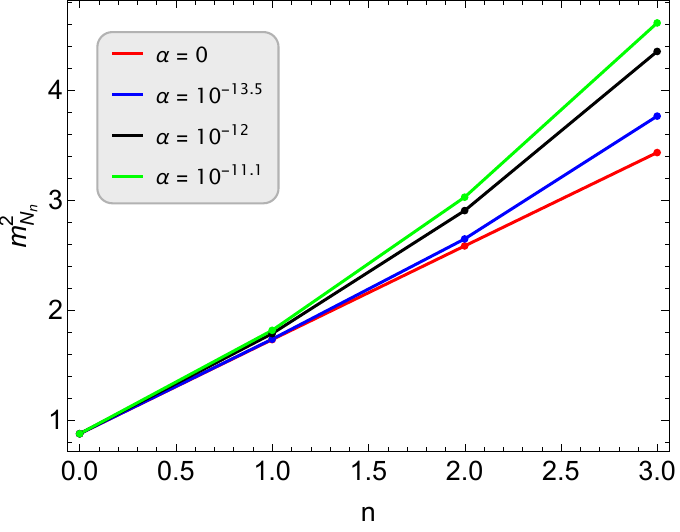}}}%
    \\ 
    {{\includegraphics[width=7.45cm]{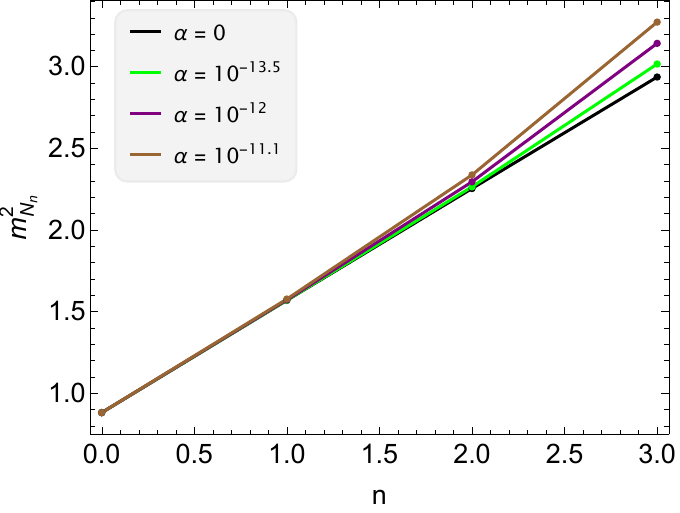}}
    \hskip 0.3cm
    {\includegraphics[width=7.2cm]{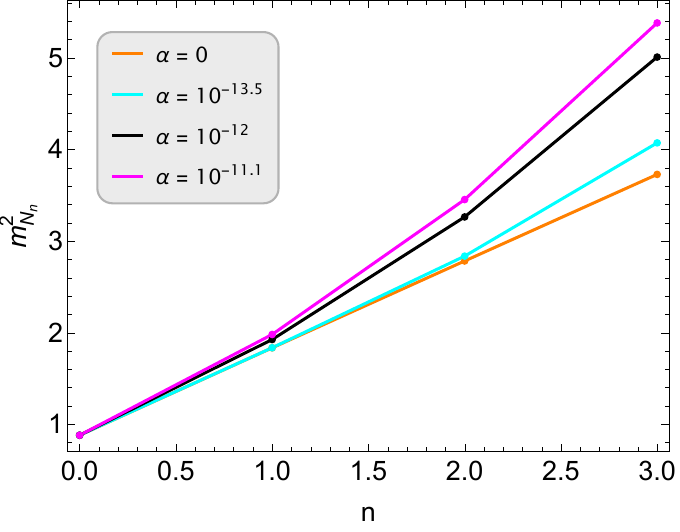}}}%
    
    \caption{Squared spectrum of nucleons in ED $(\alpha=0)$ and SD model with $\alpha = 10^{-13.5}, 10^{-12}, 10^{-11}$ as a function of $n$, considering the ground state and three radial excitations $n = 1, 2,3$. 
 {\bf Upper panels}: $\Delta = 7/2$. {\bf Lower panels}: $\Delta = 9/2$. {\bf Left panels}:  $\lambda = 0.6$.
 {\bf Right panels}: $\lambda = 0.4$.}%
    \label{Plot:Nucleonsspectrumstarobinskycoef35}%
\end{figure}

 The squared nucleon mass spectra as a function of the radial excitation number are shown in Fig.~\ref{Plot:Nucleonsspectrumstarobinskycoef35}, while the corresponding spectra are presented in Tables ~\ref {Table:NucleonsMassesm52starobinskylambda06} and \ref{Table:NucleonsMassesm52starobinskylambda04}. As the parameter $\alpha$ increases, the nucleon spectrum progressively departs from linearity with respect to the radial excitation. The impact of the SD corrections is more pronounced for higher excited states, such as $m_{N^{2}}$ and $m_{N^{3}}$. Tables \ref{Table:ReggetrajectoriesStarobinskylambda06} and \ref{Table:ReggetrajectoriesStarobinskylambda04} present the parameters obtained from the phenomenological fit to the nucleon Regge trajectories shown in Fig.~\ref{Plot:Nucleonsspectrumstarobinskycoef35}. As in SD model A, the fitted exponent $\nu$ increases with $\alpha$, indicating a progressively stronger departure from linear Regge behavior.

\begin{table}[htp!]
\centering
\begin{tabular}{c|c|c|c|c}
\hline 
\hline
$\Delta$  & $\alpha = 0$ & $10^{-13.5}$ & $10^{-12}$ & $10^{- 11.1}$\\
\hline 
7/2  & $0.729\,n + 0.883$ & $0.549(n + 1.55)^{1.09}$ &  $0.394(n + 1.89)^{1.27}$ &  $0.267(n + 2.27)^{1.46}$\\
\hline\hline
9/2 &  $0.904\,n + 0.884$ & $0.549(n + 1.52)^{1.13}$ & $0.363(n + 1.93)^{1.35}$  & $0.232(n + 2.34)^{1.58}$\\
\hline\hline
\end{tabular}
\caption{Phenomenological parametrization of nucleon Regge trajectories in the Einstein- and Starobinsky-dilaton gravities for $\lambda = 0.6$.}
\label{Table:ReggetrajectoriesStarobinskylambda06}
\end{table}

\begin{table}[htp!]
\centering
\begin{tabular}{c|c|c|c|c}
\hline 
\hline
$\Delta$  & $\alpha = 0$ & $10^{-13.5}$ & $10^{-12}$ & $10^{- 11.1}$\\
\hline 
7/2  & $0.849\,n + 0.884$ & $0.416(n + 1.71)^{1.42}$ & $0.192(n + 2.26)^{1.88}$  &  $0.157(n + 2.36)^{2.01}$\\
\hline\hline
9/2 &  $0.946\,n + 0.886$ & $0.499(n + 1.52)^{1.39}$ & $0.230(n + 2.03)^{1.91}$ & $0.222(n + 2.01)^{1.98}$\\
\hline\hline
\end{tabular}
\caption{Phenomenological parametrization of nucleon Regge trajectories in the Einstein and Starobinsky-dilaton gravities for $\lambda = 0.4$.}
\label{Table:ReggetrajectoriesStarobinskylambda04}
\end{table}


\subsection{Wave functions in SD model B}

\begin{figure}[htp!]%
    \centering
    {{\includegraphics[width=7.2cm]{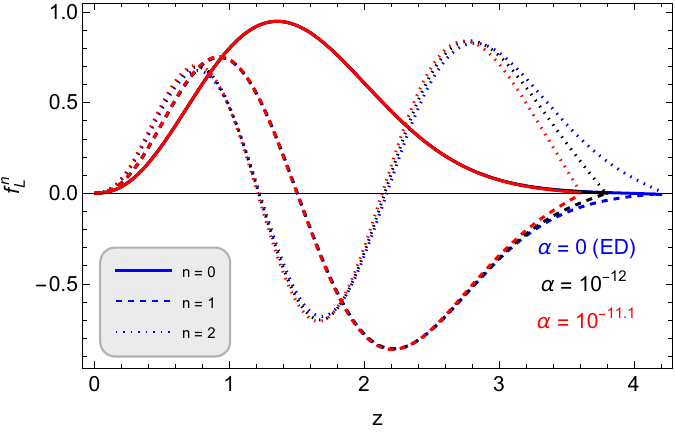}}
    \hskip 0.3cm 
    {\includegraphics[width=7.2cm]{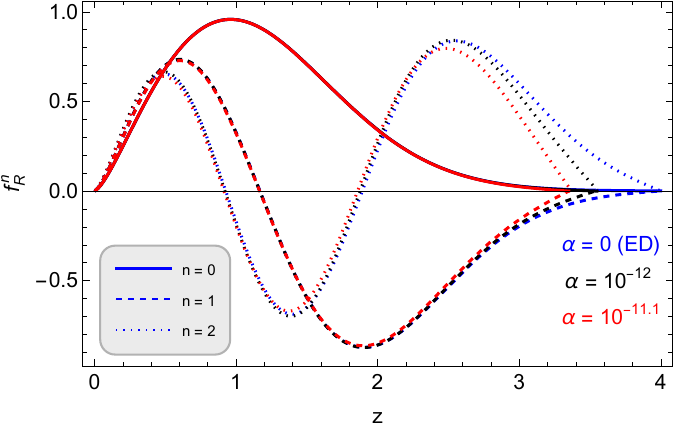}}}%
    \\ 
    \centering
    {{\includegraphics[width=7.2cm]{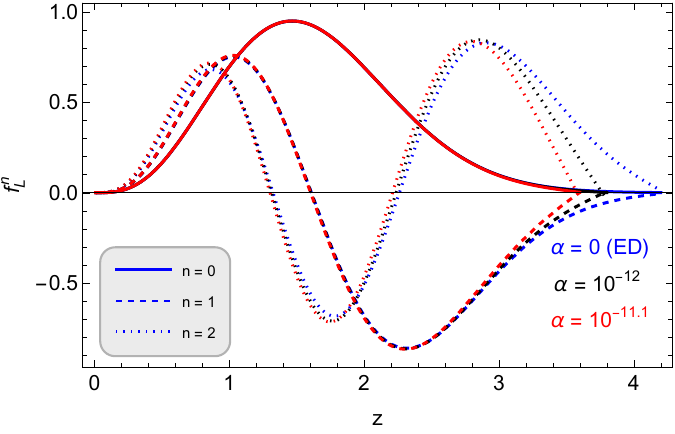}}
    \hskip 0.3cm 
    {\includegraphics[width=7.2cm]{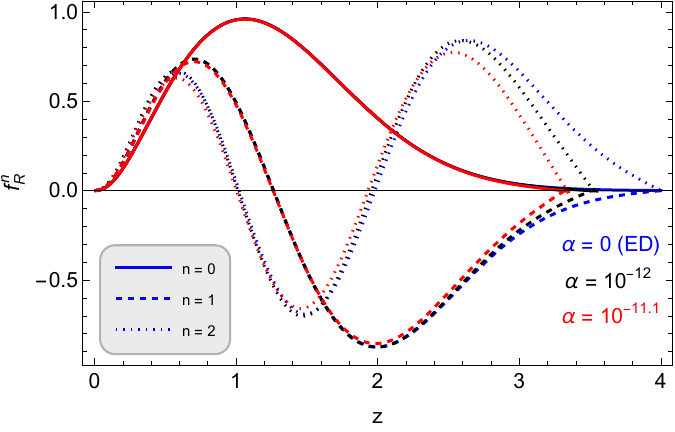}}}%
    
    \caption{Nucleon wave functions in ED $(\alpha=0)$ and SD $(\alpha=10^{-12}, 10^{-11.1})$ holographic models with $n = 0, 1, 2$  for $\lambda = 0.6$, considering $\Delta = 7/2, 9/2$.
    {\bf Upper panels}: $\Delta = 7/2$. 
 {\bf Lower panels}: $\Delta = 9/2$. {\bf Left panels}: $f_{L}^{n}(z,\alpha)$. 
 {\bf Right panels}: $f_{R}^{n}(z,\alpha)$.}%
    \label{Plot:WaveNucleonsm52coef35}%
\end{figure}

The nucleon wave functions for the ED and SD models, for $\lambda = 0.6$ considering $\alpha = 0$, $\alpha = 10^{-12}$ and $\alpha = 10^{-11.1}$ are presented in Fig.~\ref{Plot:WaveNucleonsm52coef35}, for $\Delta = 7/2$ and $9/2$. We consider the ground state, $n = 0$, and the first and second excited states of the nucleons, $n=1$, $n=2$. As $\alpha$ increases, small deviations in amplitudes emerge.

\begin{figure}[htp!]%
    \centering
    {{\includegraphics[width=7.2cm]{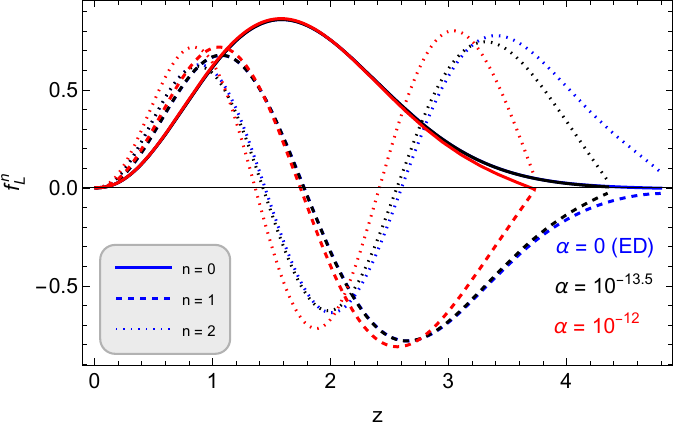}}
    \hskip 0.3cm 
    {\includegraphics[width=7.2cm]{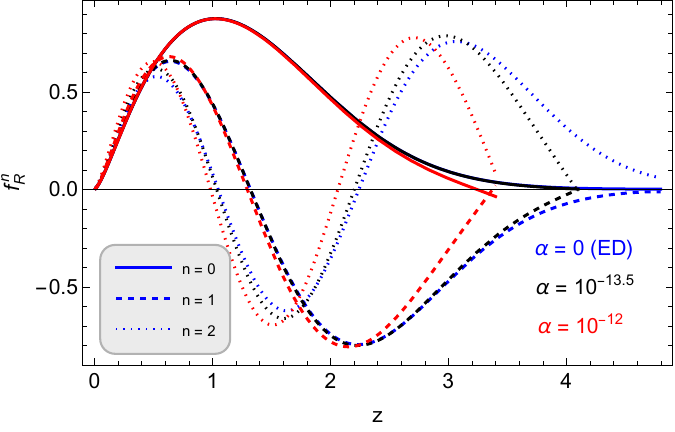}}}%
    \\ 
    \centering
    {{\includegraphics[width=7.2cm]{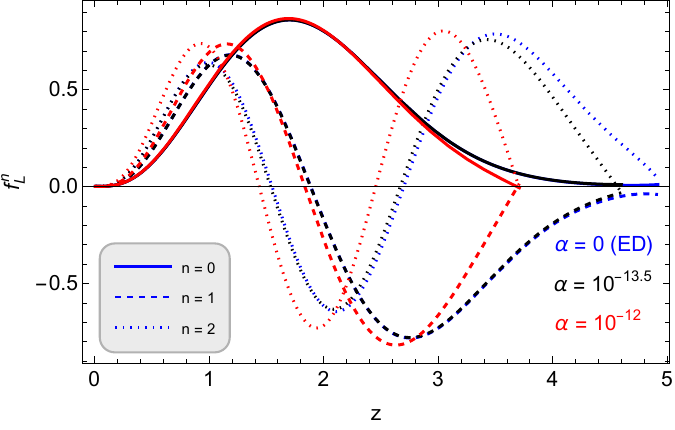}}
    \hskip 0.3cm 
    {\includegraphics[width=7.2cm]{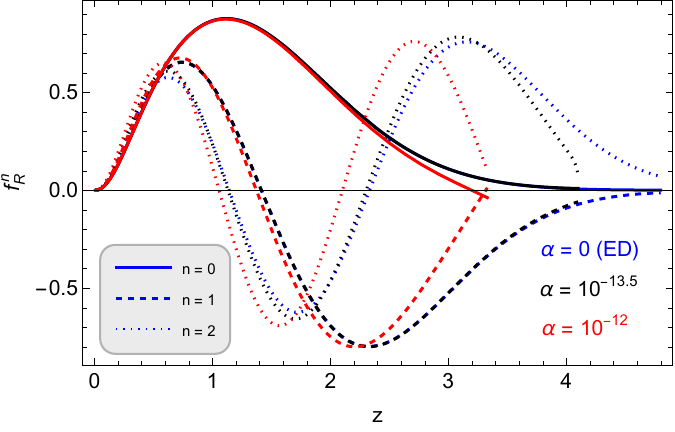}}}%
    
    \caption{Nucleon wave functions  in Einstein-dilaton $(\alpha=0)$ and Starobinsky-dilaton $(\alpha=10^{-12}, 10^{-11.1})$ holographic models with $n = 0, 1, 2$ for $\lambda = 0.4$, considering $\Delta = 7/2, 9/2$.
    {\bf Upper panels}: $\Delta = 7/2$. 
 {\bf Lower panels}: $\Delta = 9/2$. {\bf Left panels}: $f_{L}^{n}(z,\alpha)$. 
 {\bf Right panels}: $f_{R}^{n}(z,\alpha)$.}%
    \label{Plot:WaveNucleonsm52coef25}%
\end{figure}

\section{Conclusions}\label{conclusions}

        In this work, we have investigated the nucleon spectrum within the ED and SD gravity frameworks and compare our results with experimental data and the soft-wall model. The models were characterized by three (ED) or four (SD) parameters. The common three are $k$, associated with conformal symmetry breaking; conformal dimension, $\Delta$, related to the operator in dual theory, and $\lambda$, which enters the effective mass term in the nucleon Schrödinger potential. In the SD model context, additionally one has the extra parameter $\alpha$ encoding the Starobinsky correction.
 
The nucleon spectrum in both ED and SD gravity exhibits a stronger dependence on the parameters $\Delta$ and $\lambda$. In particular, in the SD model, the inclusion of $\alpha$-dependent corrections systematically shifts the nucleon spectrum toward larger masses, with the effect becoming more pronounced for the higher excited states.

The agreement with the experimental data depends on the parameters $\lambda$, $\Delta$, and $\alpha$. Within the ED framework, some choices of $\lambda$ and $\Delta$ provide a good description of the nucleon spectrum, while others exhibit noticeable discrepancies with the experimental data. In the latter case, the inclusion of SD corrections leads to an improvement of specific excited states. In particular, the $\alpha$-dependent corrections predominantly affect the higher excitations, shifting their masses toward larger values and, in some cases, bringing the theoretical predictions closer to the experimental data. Among the holographic models considered in this work, the soft-wall model exhibits the largest discrepancies with respect to the experimental data. On the other hand, when the ED results are already in good agreement with the data, sufficiently large values of $\alpha$ may lead to an overestimation of the masses, especially for the highest excited states considered.

Among the ED and SD gravities, the ED with $\lambda = 0.4$ provides the best overall agreement with the experimental nucleon spectrum.

It is worthy to mention that in the present work, we compute the nucleon spectrum associated with radial excitations. A natural extension of this study would be to analyze the nucleon spectrum as a function of spin excitations, following the approach of Ref.~\cite{deTeramond:2005su}. It would also be interesting to investigate whether it is possible to describe the linear spectrum of nucleons in SD gravity. As we have seen, the spectrum becomes non-linear for the more excited states. It would also be interesting to explore the impact of alternative $f(R)$ functions on the nucleon spectrum, such as $f(R) = R^{1 + \epsilon}$, as considered in \cite{Pretel:2022plg}. These works will be developed in the near future.

\acknowledgments

A.S.S.Jr acknowledges support from “Fundação Carlos Chagas Filho de Amparo
à Pesquisa do Estado do Rio de Janeiro” – FAPERJ, Processo SEI-260003/013507/2025. JMZP acknowledges the financial support provided by FAPERJ under Process No.~SEI-260003/000308/2024. This work is supported in part by Coordenação de Aperfeiçoamento de Pessoal de Nível Superior (CAPES) under finance code 0001. HBF is partially supported by Conselho Nacional de Desenvolvimento Cient\'{\i}fico e Tecnol\'{o}gico (CNPq) under grant  310346/2023-1, and Fundação Carlos Chagas Filho de Amparo à Pesquisa do Estado do Rio de Janeiro (FAPERJ) under grant E-26/204.095/2024.


\appendix

\bibliographystyle{utphys}

\bibliography{NucleonsStarobinsky}

\end{document}